\documentclass[pra,showpacs,superscriptaddress,amssymb,nofootinbib,twocolumn,longbibliography]{revtex4-1}
\usepackage{graphicx}
\usepackage{amsmath}
\usepackage{amsthm}
\usepackage{amssymb}
\usepackage[usenames]{color}
\usepackage{hyperref}
\usepackage{listings}
\usepackage[inline]{enumitem}

\hypersetup{
    colorlinks=true,       % false: boxed links; true: colored links=
    linkcolor=cyan,          % color of internal links
    citecolor=magenta,        % color of links to bibliography
    filecolor=magenta,      % color of file links
    urlcolor=blue,           % color of external links
    runcolor=cyan
}

\usepackage{bm}
\usepackage{threeparttable}
\usepackage{subfigure}
\usepackage{upgreek }
\usepackage{marginnote}
\usepackage{braket}
\newcommand{\beq}{\begin{equation}}
\newcommand{\eeq}{\end{equation}}
\newcommand{\beqnn}{\begin{equation*}}
\newcommand{\eeqnn}{\end{equation*}}
\newcommand{\bea}{\begin{eqnarray}}
\newcommand{\eea}{\end{eqnarray}}
\newcommand{\beann}{\begin{eqnarray*}}
\newcommand{\eeann}{\end{eqnarray*}}
\newcommand{\bes} {\begin{subequations}}
\newcommand{\ees} {\end{subequations}}

\newcommand{\abs}[1]{\ensuremath{\left| #1 \right|}}

\newtheorem{theorem}{Theorem}

\def\pr{\prime}

\def\bigo{\mathcal{O}}

\begin{document}
%%title should be as concise and catchy as possible; "analytical approximation" doesn't add much in that regard
%\title{Analytic approximations to quasi-adiabatic Grover search}
\title{Quasi-adiabatic Grover search via the WKB approximation}
\author{Siddharth Muthukrishnan}\email{muthukri@usc.edu}
\affiliation{Department of Physics and Astronomy, University of Southern California, Los Angeles, California 90089, USA}
\affiliation{Center for Quantum Information Science \& Technology, University of Southern California, Los Angeles, California 90089, USA}

\author{Daniel A. Lidar}
\affiliation{Department of Physics and Astronomy, University of Southern California, Los Angeles, California 90089, USA}
\affiliation{Center for Quantum Information Science \& Technology, University of Southern California, Los Angeles, California 90089, USA}
\affiliation{Department of Electrical Engineering, University of Southern California, Los Angeles, California 90089, USA}
\affiliation{Department of Chemistry, University of Southern California, Los Angeles, California 90089, USA}

\begin{abstract}
In various applications one is interested in quantum dynamics at intermediate evolution times, for which the adiabatic approximation is inadequate. Here we develop a quasi-adiabatic approximation based on the WKB method, designed to work for such intermediate evolution times. We apply it to the problem of a single qubit in a time-varying magnetic field, and to the Hamiltonian Grover search problem, and show that already at first order, the quasi-adiabatic WKB captures subtle features of the dynamics that are  missed by the adiabatic approximation. However, we also find that the method is sensitive to the type of interpolation schedule used in the Grover problem, and can give rise to nonsensical results for the wrong schedule. Conversely, it reproduces the quadratic Grover speedup when the well-known optimal schedule is used.

\end{abstract}

\maketitle

\section{Introduction}

There exist only a handful of Hamiltonian-based quantum algorithms \cite{FarhiAnalog,Farhi:00}, designed to run on analog quantum computers \cite{Feynman:1985ul,Peres:1985aa,Margolus:90}, that exhibit a provable quantum speedup \cite{Albash-Lidar:RMP}. The adiabatic version of the Grover search problem is one such example \cite{Roland:2002ul}. 
%I.e., while there is a formal equivalence between the quantum circuit model and the adiabatic model~\cite{aharonov_adiabatic_2007}, translating a quantum algorithm from the circuit model to the adiabatic using the formal equivalence yields unwieldy Hamiltonians in the adiabatic paradigm.
The existence of this speedup is proven using the adiabatic theorem \cite{Jansen:07}, i.e., it is based on an \emph{asymptotic} analysis in the total evolution time. 
%The analysis further makes use of the fact that the evolution is effectively restricted to a two-dimensional subspace.
This is in contrast to the circuit model version of the Grover problem \cite{Grover:1996}, for which a closed-form analytical solution is known for arbitrary evolution times and arbitrary initial amplitude distributions~\cite{Biham:2000aa,Biham:1999ye}. No such closed form analytical solution of the Hamiltonian version of Grover's algorithm is known as of yet. 

The Wentzel-Kramers-Brillouin (WKB) method is a famous technique for approximating differential equations which has found applications in many domains of physics and mathematics, including optics, acoustics, astrophysics, elasticity, and quantum mechanics (see Ref.~\cite{schlissel1977initial} for a mathematical history of the WKB method). In this work, we adopt the WKB method to provide approximate analytical solutions to the Hamiltonian Grover problem. The WKB method we use is quasi-adiabatic (as opposed to semiclassical \cite{Messiah:book}): the small parameter is the inverse of the total evolution time (not $\hbar$, which we set to $1$). We choose to focus on the Grover problem since this problem is well studied and understood, but the WKB method is widely applicable and easily generalizable to other Hamiltonian-based quantum algorithms. We thus expect it to be a useful tool in analyzing such algorithms beyond the adiabatic approximation.

We compare the results of the WKB approximation with a numerically exact solution. Strikingly, we find that the quality of WKB results depends strongly on the interpolation schedule from the initial to the final Hamiltonian. The WKB approximation is reliable already at low order for the schedule that generates a quantum speedup for the Grover problem \cite{Roland:2002ul}, but fails for the other schedules we tested. These other schedules are characterized by a different dependence on the power of the inverse spectral gap. 

The structure of the paper is as follows. We briefly review the quasi-adiabatic WKB method in Sec.~\ref{sec:WKBmethod}. The method is applied to the Grover problem in Sec.~\ref{sec:Grover}, and the WKB solutions are derived in Sec.~\ref{sec:WKBmethodDetailed}. The results are discussed and analyzed in Sec.~\ref{sec:results}, where we perform a comparison with the numerically exact solution. We conclude in Sec.~\ref{sec:conc}. 

We remark that there are other tools available to study quasi-adiabatic dynamics: Adiabatic perturbation theory is a popular method~\cite{Teufel:book}. In Appendix~\ref{app:Hage} we study a particular variant of adiabatic perturbation theory from Ref.~\cite{Hagedorn:2002kx} and compare it to our method. Our method is not a variant of adiabatic perturbation theory because at the lowest order we do not recover the adiabatic evolution.

\section{Quasi-adiabatic WKB for interpolating Hamiltonians}
\label{sec:WKBmethod}

We start by briefly reviewing the asymptotic WKB expansion technique (for background see, e.g., Ref.~\cite{holmes2012introduction}), and connect it to interpolating Hamiltonians of the type used in adiabatic quantum computing. 

\subsection{WKB as an asymptotic expansion}
The WKB expansion
\beq
y(r) \sim e^{\theta(r)/\epsilon} [y_0(r) + \epsilon y_1(r) + \epsilon^2 y_2(r) +\dots] ,
\eeq
is an ansatz used for the solution of ordinary differential equations in $y(r)$ that contain a small parameter, $\epsilon$, multiplying the highest derivative. This ansatz is an asymptotic expansion in $\epsilon$, i.e., there is no guarantee that it will provide a unique or even a convergent solution. In fact, the asymptotic series for $y(r)$ is usually divergent; the general term $\epsilon^n y_n(r)$ starts to increase after a certain value $n=n_{\max}$, which can be estimated for second order differential equations of the form $\epsilon^2 y''(r)=Q(r)y(r)$, if $Q(r)$ is analytic \cite{Winitzki:2005aa}. The number  $n_{\max}$ can be interpreted as the number of oscillations between $r_0$ [the point at which $y(r)$ needs to be evaluated] and the turning point $r_{\star}$ [i.e., where $Q(r_{\star})=0]$ closest to $r_0$. In this work we will only be concerned with the expansion up to order $\epsilon$ for a second order differential equation. For later convenience, we list the expressions for the derivatives of the ansatz:
\bes \label{eq:ansatz}
\begin{align}
y &\sim e^{\theta/\epsilon} \sum_{j=0}^\infty \epsilon^{j} {y_j} \\
y^\pr &\sim e^{\theta/\epsilon} \sum_{j=0}^\infty \epsilon^{j-1}\underbrace{(\theta^\pr y_j + y^\pr_{j-1})}_{\equiv z^{(1)}_j}  \\
y^{\pr\pr} &\sim e^{\theta/\epsilon} \sum_{j=0}^\infty \epsilon^{j-2}\underbrace{[(\theta^\pr)^2 y_j + \theta^{\pr\pr} y_{j-1} + 2\theta^\pr y^\pr_{j-1} + y^{\pr\pr}_{j-2}]}_{\equiv z^{(2)}_j}
\end{align}
\ees
with $y_k \equiv 0$ if $k<0$, and where the number of primes denotes the order of the derivative. 
%The quantities $z^{(1),(2)}_j$ are defined for later convenience.

%We will use this ansatz to approximate the amplitudes of the quantum wavefunction corresponding to our Hamiltonians.

\subsection{Interpolating Hamiltonians}

We consider interpolating Hamiltonians of the form
\beq
H[r(s)]=[1-r(s)]H_\mathrm{initial} + r(s) H_\mathrm{final}.
\eeq
which depend on time only via the dimensionless time $s \equiv t/t_f$. Here $t_f$ denotes the total evolution time and is the only timescale in the problem.  The ``interpolation schedule" $r(s)$ is strictly increasing, differentiable, and satisfies the boundary conditions $r(0) = 0$ and $r(1) = 1$. The derivative of the inverse of $r(s)$, viz. $s^\pr(r)$, is therefore also strictly positive. This allows us to divide by $s^\pr$ when we need to.

Consider now the Schr{\"o}dinger equation for this evolution
\beq
i \frac{d}{d t} \ket{\chi(t)} = \mu H(r[s(t)])\ket{\chi(t)},
\eeq
where $\mu$ is an energy scale, and $H(\cdot)$ is dimensionless, e.g., a linear combination of Pauli matrices.
Writing everything in terms of $s$, we get
\beq
i \frac{d}{d s} \ket{\chi(s)} = \mu t_f H[r(s)]\ket{\chi(s)}.
\eeq
One can also write the problem in terms of $r$. This yields:
\beq 
\label{eq:rescaleSchro}
i \epsilon \frac{d}{d r} \ket{\chi(r)} = g (r)H(r)\ket{\chi(r)},
\eeq
where $g(r) \equiv s^\pr (r)$, $s(r): [0,1] \mapsto [0,1]$,
and where 
\beq
\epsilon \equiv \frac{1}{\mu t_f},
\eeq 
is the dimensionless small parameter for our WKB expansion. Since $\epsilon$ is small for large $t_f$, we call our method ``quasi-adiabatic WKB".%
\footnote{We remark that the quasi-adiabatic WKB approximation should not be confused with the traditional  WKB approximation associated with the $\hbar \to 0$ limit. The latter is typically used as a semiclassical approximation in one-dimensional position-momentum quantum mechanics, involving a potential barrier (see, e.g., Ref.~\cite{Messiah:book}). The quasi-adiabatic and semiclassical WKB approximations are not interchangeable. This can be seen from the Schr{\"o}dinger equation for a particle in a one-dimensional potential:
\[
i \frac{\hbar}{t_f} \frac{d}{d s}\ket{\chi} = \left(-\frac{\hbar^2}{2m} \partial_x^2 + V(x, s t_f) \right) \ket{\chi} ,
\]
where again $s=t/t_f$ and $V(x,t)$ is a space- and time-dependent potential energy function. It is evident that there is no way to trade both $\hbar$ and $1/t_f$ for a single small parameter, since they appear together as the product $\hbar t_f$.} 

%Therefore, the limits $\hbar \to 0$ and the $t_f\gg 1$ can be taken independently of each other and correspond to different approximation schemes.

\section{The Grover problem via the quasi-adiabatic WKB approximation}
\label{sec:Grover}

%We now turn to the specific interpolating Hamiltonians we study and their WKB approximations.

%\subsection{The Grover Hamiltonian}

Recall that the Grover problem can be formulated as finding an item in an unsorted list of $N=2^n$ items, in the smallest number of queries \cite{Grover:97a}. This admits a quadratic quantum speedup, as was first shown by Grover in the circuit model \cite{Grover:1996}. It is also one of the few instances where an adiabatic algorithm was discovered which recovers the quantum speedup. The crucial insight, which eluded the first attempt \cite{Farhi:00}, was that the speedup obtained in the circuit model could be recovered in the adiabatic model provided the right interpolation schedule $r(s)$ is chosen, namely, a schedule that drives the system more slowly when the gap is smaller \cite{Roland:2002ul} (see also Refs.~\cite{Jansen:07,RPL:10}). 

In the Hamiltonian Grover algorithm one uses the $n$-qubit interpolating Hamiltonian
\beq
H_\mathrm{Grover}[r(s)] = [1-r(s)] (I-\ket{u}\bra{u}) + r(s)(I-\ket{m}\bra{m}),
\eeq
where $m \in \{0,1\}^n$ is the marked state and
\beq
\ket{u} \equiv \frac{1}{\sqrt{2^n}} \sum_{x \in \{0,1\}^n} \ket{x},
\eeq
is the uniform superposition state. The system is initialized in the state $\ket{u}$.
It can be easily checked that the dynamics described by this Hamiltonian is restricted to $\mathcal{S} \equiv \mathrm{Span}\{\ket{u},\ket{m}\}$. Let $K \equiv 2^n-1$ 
and define
\beq
\ket{m^\perp} \equiv \frac{1}{\sqrt{K}} \sum_{\substack{x \in \{0,1\}^n \\ x \neq m}} \ket{x},
\eeq
so that $\ket{u} = (\ket{m}+\sqrt{K}\ket{m^\perp})/\sqrt{K+1}$.
Note that $\{\ket{m}, \ket{m^\perp}\}$ is an orthonormal basis for $\mathcal{S}$. In this basis, the Hamiltonian is
\beq \label{eq:groverRestrictHam}
H (s) = \begin{pmatrix} [1-r(s)] \frac{K}{K+1} & -[1-r(s)] \frac{\sqrt{K}}{K+1} \\ -[1-r(s)] \frac{\sqrt{K}}{K+1} & 1-[1-r(s)] \frac{K}{K+1} \end{pmatrix}.
\eeq
Henceforth we restrict our analysis to this two-dimensional Hamiltonian and will not return to the high-dimensional Hamiltonian that gave rise to it.

Let 
\beq
\ket{\chi(s)}=\psi(s)\ket{m}+\phi(s)\ket{m^\perp}\ ,
\eeq
i.e., henceforth $\psi(s)$ is the amplitude of the marked state (ground state of the final Hamiltonian), and $\phi(s)$ is the amplitude of the unmarked component (the excited state of the final Hamiltonian). 

From Eq.~\eqref{eq:rescaleSchro}, the Schr{\"o}dinger equation for a general interpolation becomes
\bes
\label{eq:diffGrover}
\begin{align}
i \epsilon \psi^\pr &= \frac{g(r)}{K+1} \left[ K(1-r) \psi - \sqrt{K} (1-r) \phi \right] ,\label{eq:diffGrover1}\\
i \epsilon \phi^\pr &= \frac{g(r)}{K+1} \left[ -\sqrt{K}(1-r) \psi + (1+rK) \phi \right] . \label{eq:diffGrover2}
\end{align}
\ees
The boundary conditions are
\beq
 \psi(0)=\frac{1}{\sqrt{K+1}},\quad \phi(0)=\sqrt{\frac{K}{K+1}},
\eeq
so it follows from Eqs.~\eqref{eq:diffGrover} that $\psi^\pr(0)=\phi^\pr(0)=0$.

We now turn the above coupled first order system into two decoupled second order differential equations:
\bes
\begin{align}
\label{eq:upGrover2nd}
&\epsilon^2 (1-r)\psi''+\epsilon (\epsilon a_{1,1}+a_{1,2})\psi'+a_0\psi=0 \\
 \label{eq:downGrover2nd}
 &\epsilon^2 (1-r)\phi''+\epsilon (\epsilon a_{1,1}+a_{1,2})\phi'+(a_0 + ig\epsilon)\phi=0\ ,
\end{align}
\ees
where
\bes
\begin{align}
a_0 &= - \frac{g^2 K (1-r)^2 r}{K+1}\\
a_{1,1}&= 1-\frac{g^\pr}{g}(1-r),\quad a_{1,2}= i  (1-r) g\ .
\end{align}
\ees

The function $g(r)=s'(r)$ uniquely determines the schedule $r(s)$. We shall consider four different schedules corresponding to choices $\alpha\in\{0,1,2,3\}$ in
\beq \label{eq:schedgen}
r'(s) = c_\alpha \Delta(r)^\alpha\ ,
\eeq
where $c_\alpha$ is a constant that depends on $K$ (see Refs.~\cite{Jansen:07,Roland:2002ul}) and $\Delta(r)$ is the eigenvalue gap of the Hamiltonian in Eq.~\eqref{eq:groverRestrictHam}, given by:
\beq
\Delta(r) = \sqrt{1-\frac{4K r(1-r)}{K+1}}.
\label{eq:gap}
\eeq
Equation~\eqref{eq:schedgen} forces the schedule to become slower (faster) when the gap is smaller (larger). 

The linear schedule [$r(s)=s$] corresponds to the choice $\alpha=0$, and the schedule discovered by Roland and Cerf \cite{Roland:2002ul} corresponds to $\alpha=2$. We also analyze schedules corresponding to $\alpha=1,3$. To find the constant $c_\alpha$, we integrate Eq.~\eqref{eq:schedgen} and use the boundary condition $s(1)=1$. The expressions for the schedules thus obtained, expressed 
in terms of the corresponding $g_\alpha(r)$ functions [recall that $g(r) \equiv s^\pr (r)$], are as follows:
\bes
\label{eq:4schedsprime}
\begin{align}
g_0(r) &= g_\mathrm{lin}(r) =1 \\
g_1(r) &= \frac{2\sqrt{\frac{K}{K+1}}}{\log \left(\frac{\sqrt{K+1}+\sqrt{K}}{\sqrt{K+1}-\sqrt{K}}\right)} \times \frac{1}{\Delta(r)}\\
g_2(r) &= g_\mathrm{RC}(r) = \frac{\sqrt{K}}{(K+1)\tan^{-1}(\sqrt{K})} \times \frac{1}{\Delta(r)^2} \label{eq:optsprime} \\
g_3(r) &=\frac{1}{K+1} \times \frac{1}{\Delta(r)^3}
\end{align}
\ees

We now turn to the construction of the WKB solutions for both amplitudes $\psi$ and $\phi$ for each of the schedules. 

\section{Constructing the WKB solutions}
\label{sec:WKBmethodDetailed}

To derive the WKB solutions, we substitute the WKB ansatz [Eqs.~\eqref{eq:ansatz}] into Eqs.~\eqref{eq:upGrover2nd} and \eqref{eq:downGrover2nd}.\footnote{A Mathematica\textsuperscript{\textregistered} notebook containing code for obtaining the WKB expressions used in our analysis is provided at \url{https://tinyurl.com/WKB-notebook}.
} Then, we set the terms multiplying different orders $\epsilon^j$ to zero, which yields the following recursive set of equations for $j\geq 1$:

\bes
\label{eq:21}
\begin{align}
\label{eq:21a}
&(1-r)z_j^{(2)}+a_{1,1}z_{j-1}^{(1)}+a_{1,2} z_j^{(1)} +a_0 y_j= 0 \\
\label{eq:21b}
&(1-r)z_j^{(2)}+a_{1,1}z_{j-1}^{(1)}+a_{1,2} z_j^{(1)} +a_0 y_j+ i g y_{j-1}= 0\ ,
\end{align}
\ees
where $\psi$ [Eq.~\eqref{eq:upGrover2nd}] is reconstructed from Eq.~\eqref{eq:21a}, and $\phi$ [Eq.~\eqref{eq:downGrover2nd}] is reconstructed from Eq.~\eqref{eq:21b}.
We consider only the lowest  three orders in $\epsilon$ below.

First, isolating the $\epsilon^0$ term [i.e., setting $j=0$ in both Eqs.~\eqref{eq:21a} and \eqref{eq:21b}], we obtain the \emph{eikonal} equation:
\beq
(\theta^\pr)^2 + ig \theta^\pr - g^2 \frac{K r(1-r)}{K+1} = 0\ ,
\label{eq:eikonal}
\eeq
which is a quadratic equation in $\theta'$, so that:
\beq
\theta_{\pm}^\pr =  -\frac{i g}{2} \left[1 \pm \Delta(r) \right]\ .
\label{eq:23}
\eeq

Turning to the $\epsilon^1$ term, we obtain the \emph{transport} equations:
\bes
\label{eq:transportGroverGeneral}
\begin{align} 
\label{eq:25a}
\frac{\psi_0^\pr}{\psi_0} &= -\frac{(1-r) \theta^{\pr \pr} + \left[1-\frac{g^\pr}{g}(1-r)\right]  \theta^\pr}{(1-r)(2  \theta^\pr + ig)}\ , \\
\label{eq:25b}
\frac{\phi_0^\pr}{\phi_0} &= -\frac{(1-r) \theta^{\pr \pr} + \left[1-\frac{g^\pr}{g}(1-r)\right]  \theta^\pr + i g}{(1-r)(2  \theta^\pr + ig)}\ .
\end{align}
\ees
Here $\psi_0,\phi_0$ are the parts of the WKB approximant that correspond to $y_0$, which was a general placeholder for the lowest order term. Further,  Eq.~\eqref{eq:25a} is obtained from Eq.~\eqref{eq:21a}, and Eq.~\eqref{eq:25b} is obtained from Eq.~\eqref{eq:21b}, both after setting $j=1$ and using the eikonal equation~\eqref{eq:eikonal} to eliminate the $y_1$ term.
Let $\Theta_{\pm} \equiv \theta_{\pm}^\pr/g(r)$.
It is easy to check that the transport equations then become:
\bes
\label{eq:transportGroverGeneral2}
\begin{align} 
\label{eq:26a}
\frac{\psi_0^\pr}{\psi_0} &= -\frac{(1-r) \Theta^{\pr} + \Theta}{(1-r)(2  \Theta + i)}\ , \\
\label{eq:26b}
\frac{\phi_0^\pr}{\phi_0} &= -\frac{(1-r) \Theta^{\pr} +  \Theta + i }{(1-r)(2  \Theta + i)}\ .
\end{align}
\ees
Since, by Eq.~\eqref{eq:23}, $\Theta_{\pm} = -\frac{i}{2} \left[1 \pm \Delta(r) \right]$ is independent of $g$, it follows that $\psi_0$ and $\phi_0$ do not depend on the interpolation $g$.

Further, using $\Theta_+=-(i+\Theta_-)$ and $\Theta'_\pm=\mp\frac{i}{2}\Delta'$, it is straightforward to show that the r.h.s. of Eq.~\eqref{eq:26a} for $\psi_0^\pm$ is identical to the r.h.s. of Eq.~\eqref{eq:26b} for $\phi_0^\mp$. Thus, after integration we have $\psi_0^\pm(r)=c_0^\pm\phi_0^\mp(r)$, where $c_0^\pm$ is (the exponential of) an integration constant. 

Moreover, the r.h.s. of Eq.~\eqref{eq:26a} corresponding to $\Theta_{\pm}$ is easily seen to be equal to $
-\frac{1}{2}\left[\frac{\Delta'(r)}{\Delta (r)}+\frac{1}{1-r}\pm\frac{1}{(1-r) \Delta (r)}\right]$. Hence, integrating Eqs.~\eqref{eq:transportGroverGeneral2} yields:
\begin{align} 
\label{eq:27}
\log\psi_0^\pm(r) &= \log \phi_0^\mp(r) + \tilde{c}_0^\pm  \\
&= \frac{1}{2}\log {\frac{1-r}{\Delta(r)}}\mp\frac{1}{2} \int \frac{1}{(1-r) \Delta (r)} \, dr + \tilde{d}_0^\pm\ ,\notag
\end{align}
where $\tilde{c}_0^\pm$ and $\tilde{d}_0^\pm$ are integration constants. Or, using the explicit form for the gap given in Eq.~\eqref{eq:gap}:
\bes
\label{eq:28}
\begin{align} 
\psi_0^+(r) &= c_0^+\phi_0^-(r)  \\
&= d_0^+\frac{1-r}{\sqrt{\sqrt{K+1}\Delta(r)[K (2r-1)+(K+1) \Delta (r)+1]}} \, \notag
\label{eq:28a} \\
\psi_0^-(r) &= c_0^-\phi_0^+(r) \\
&= d_0^-\sqrt{\frac{K (2r-1)+(K+1) \Delta (r)+1}{\sqrt{K+1}\Delta(r)}}  \ ,\notag
\label{eq:28b}
\end{align}
\ees
where $c_0^\pm=e^{\tilde{c}_0^\pm}$ and $d_0^\pm=e^{\tilde{d}_0^\pm}$.

Finally, turning to the $\epsilon^2$ term [i.e., setting $j=2$ in Eqs.~\eqref{eq:21a} and \eqref{eq:21b}] yields:
\beq
w^\pr = - \frac{ y_0^{\pr\pr}(1-r)+ \left[1- \frac{g^\pr}{g}(1-r) \right] y_0^\pr }{(1-r) (2\theta^\pr + i g)y_0}\ ,
\label{eq:w'}
\eeq
where $w \equiv \frac{y_1}{y_0}$. Here $y$ represents both $\psi$ and $\phi$. We used the eikonal equation to eliminate the $y_2$ term, and the transport equations to obtain $y_0'/y_0$ in $w'$.
Solving Eq.~\eqref{eq:w'} yields $y_1$. Note that here we cannot remove the dependence of $y_1$ on the interpolation $g$. 

We can now assemble the different functions into a solution. Given the interpolation $g$, we can integrate Eq.~\eqref{eq:23} to find $\theta_\pm$, resulting in two solutions $\psi^\pm$ and $\phi^\pm$. This means that we have to consider linear combinations of these two solutions. Thus
\bes
\label{eq:WKBgeneralsols}
\begin{align}
\psi &\sim A_{\psi} e^{\theta_+/\epsilon} (\psi_0^+ + \epsilon \psi_1^+) + B_{\psi} e^{\theta_-/\epsilon} (\psi_0^- + \epsilon \psi_1^-) \\
\phi &\sim A_{\phi} e^{\theta_+/\epsilon} (\phi_0^+ + \epsilon \phi_1^+) + B_{\phi} e^{\theta_-/\epsilon} (\phi_0^- + \epsilon \phi_1^-)\ ,
\end{align}
\ees
where the constants $A_{\psi,\phi},B_{\psi,\phi}$ are determined using the boundary conditions $\psi(0)=\frac{1}{\sqrt{K+1}}$, $\phi(0)=\sqrt{\frac{K}{K+1}}$,  and $\psi^\pr(0)=\phi^\pr(0)=0$. Note that despite the fact that $\psi_0$ and $\phi_0$ do not depend on $g$, the parameter $\theta$ does, via $\theta_\pm = \int g \Theta_\pm dr$. Therefore even at the lowest order, the approximate solution retains a dependence on the interpolation $g$.  

The only constraints our solutions must satisfy are the differential equations~\eqref{eq:21} and the boundary conditions. Thus we are free to choose the integration constants ($c_0^\pm,d_0^\pm$, and others that would arise at higher orders $j\geq 2$), and henceforth we choose them to be equal at all orders, such that only $A,B,C,D$ are undetermined until we use the boundary conditions. 

It is important to remember that the WKB approximation method does not enforce normalization. Hence, generically, the WKB approximation to a quantum state is unnormalized, resulting in approximations to probabilities that may be greater than $1$.\footnote{For convenience, we will abuse terminology somewhat and refer to the WKB approximations to physical probabilities as ``probabilities" even though they may not be normalized. It should be clear from the context whether we are referring to approximated probabilities or to actual probabilities.} Thus, care must be taken when applying this approximation technique to estimate physical quantities, and in particular one must check that normalization holds. For some of the examples we study here, such nonsensical probabilities indeed arise. In Sec.~\ref{sec:n-Grover}, we study whether the norm of the WKB approximation is an indicator of approximation quality and also whether renormalization can be improve the WKB approximation.
 
One final general comment is in order. It turns out that the differential equations \eqref{eq:upGrover2nd} and \eqref{eq:downGrover2nd} have the following unfortunate property: substituting the WKB approximation to $\psi$ into Eq.~\eqref{eq:diffGrover1} and solving for $\phi$ does {not} yield a good approximation to $\phi$. On other hand, the WKB approximation to $\phi$ does yield a good approximation to $\phi$. This is why we need to perform the WKB approximation separately for each of the amplitudes.

\section{Results}
\label{sec:results}

In this section we analyze the quality of the approximate solutions by comparing them with the solutions obtained via numerical integration of the Schr{\"o}dinger equation. The results obtained by numerical integration are sufficiently accurate that we can take the numerical solution to be a good proxy to the exact solution. We denote the numerically obtained solution by $\ket{\chi_\mathrm{Num}}$ and the solution obtained from the WKB approximation by $\ket{\chi_\mathrm{WKB}}$.

%%Three figures below
%
\begin{figure*}[th]%Population dynamics with s
%\hspace*{\fill}
\subfigure[]{\includegraphics[width = 0.97\columnwidth]{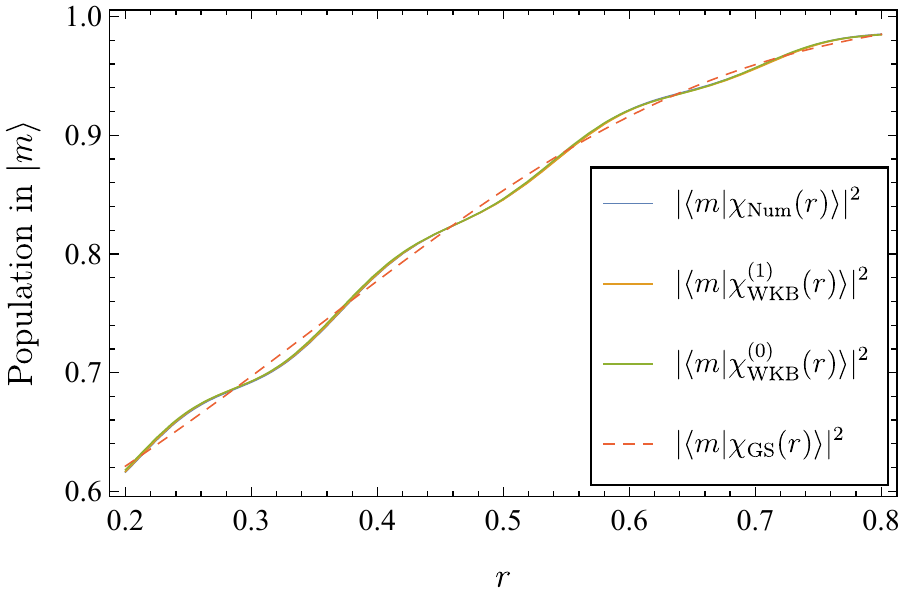}\label{fig:comparetf50}}\hfill
\subfigure[]{\includegraphics[width = \columnwidth]{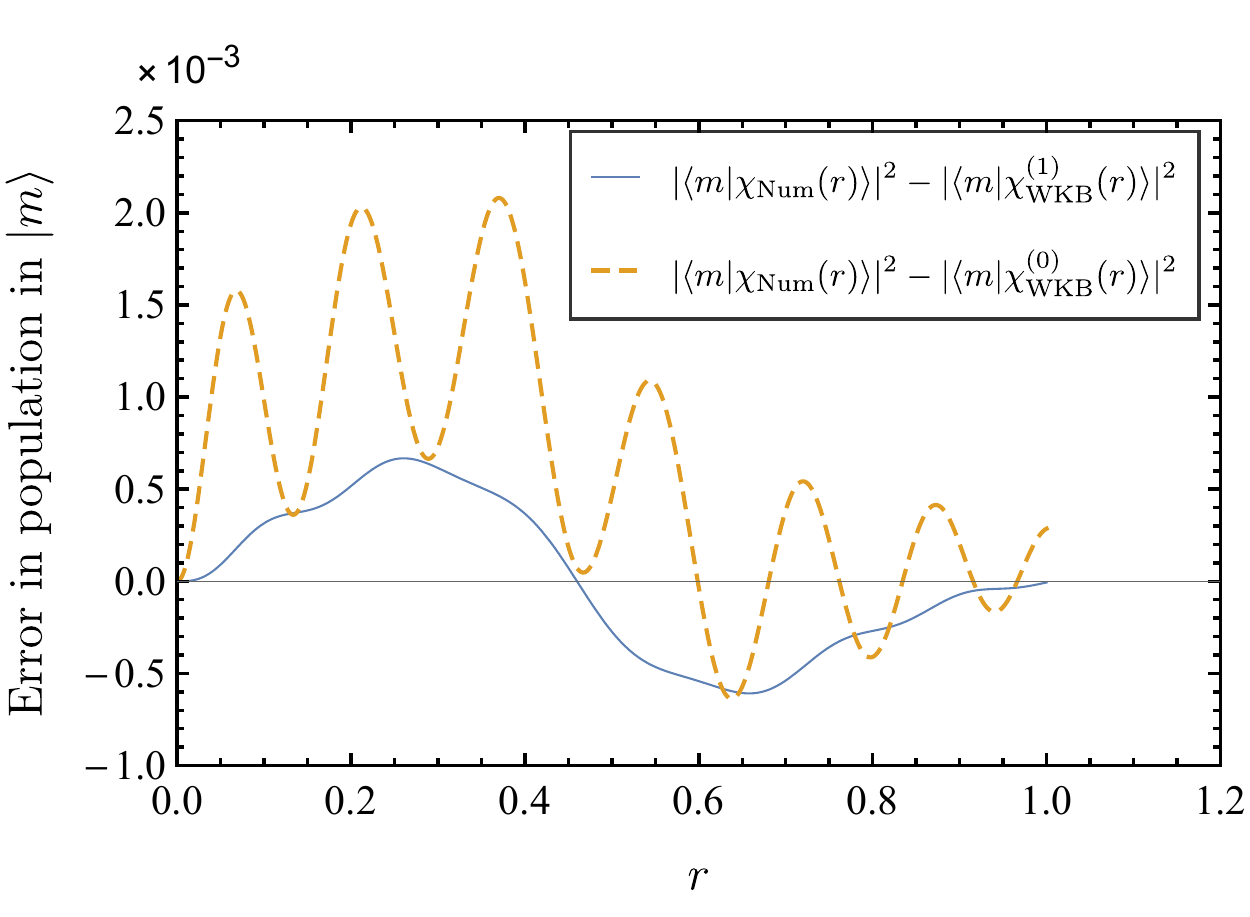}\label{fig:errortf50}}
%\hspace*{\fill}
\caption{Evolution of a single qubit ($n=1$) in a magnetic field, under the $g_0(r)=1$ schedule. (a) Population in the ground state $\ket{m}$ as a function of $r=s=t/t_f$ for $t_f=50$ according to the numerical solution ($\ket{\chi_\mathrm{Num}}$), the two lowest orders of the WKB approximation ($\ket{\chi^{(0)}_\mathrm{WKB}}$ and $\ket{\chi^{(1)}_\mathrm{WKB}}$), and the naive adiabatic evolution ($\ket{\chi_\mathrm{GS}}$). The WKB predictions and the numerical solution exhibit oscillations and are indistinguishable from each other on the scale shown. The adiabatic approximation does not exhibit oscillations. (b) The ground state population difference between the WKB approximation and the numerical simulation for $t_f=50$. The higher order WKB approximation provides a significantly better approximation.}
\label{fig:popDynamics_n=1}
\end{figure*}
\begin{figure*}[!htbp]%Population dynamics with t_f
%\hspace*{\fill}
\subfigure[]{\includegraphics[width = 0.97\columnwidth]{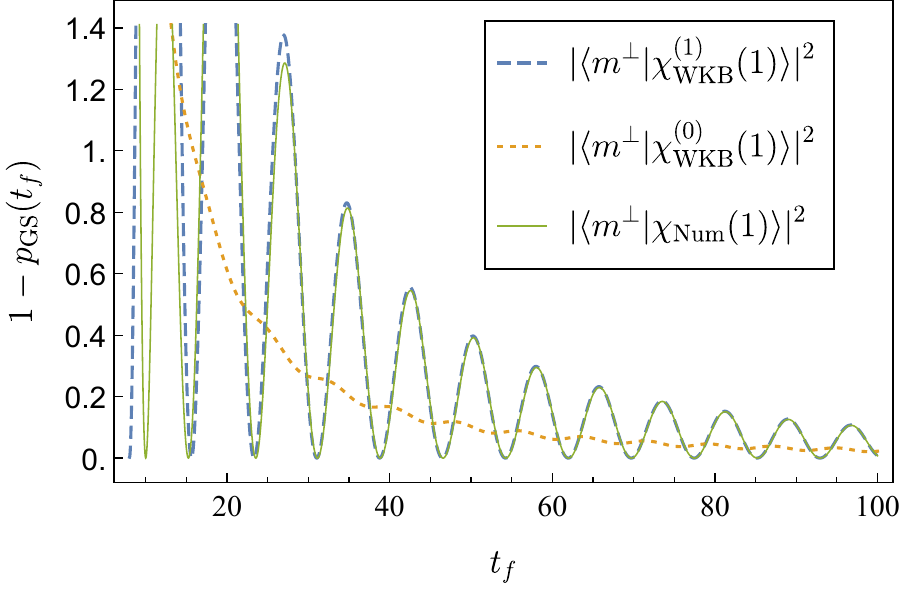}\label{fig:2ordersWKBfinalpop}} \hfill
\subfigure[]{\includegraphics[width = \columnwidth]{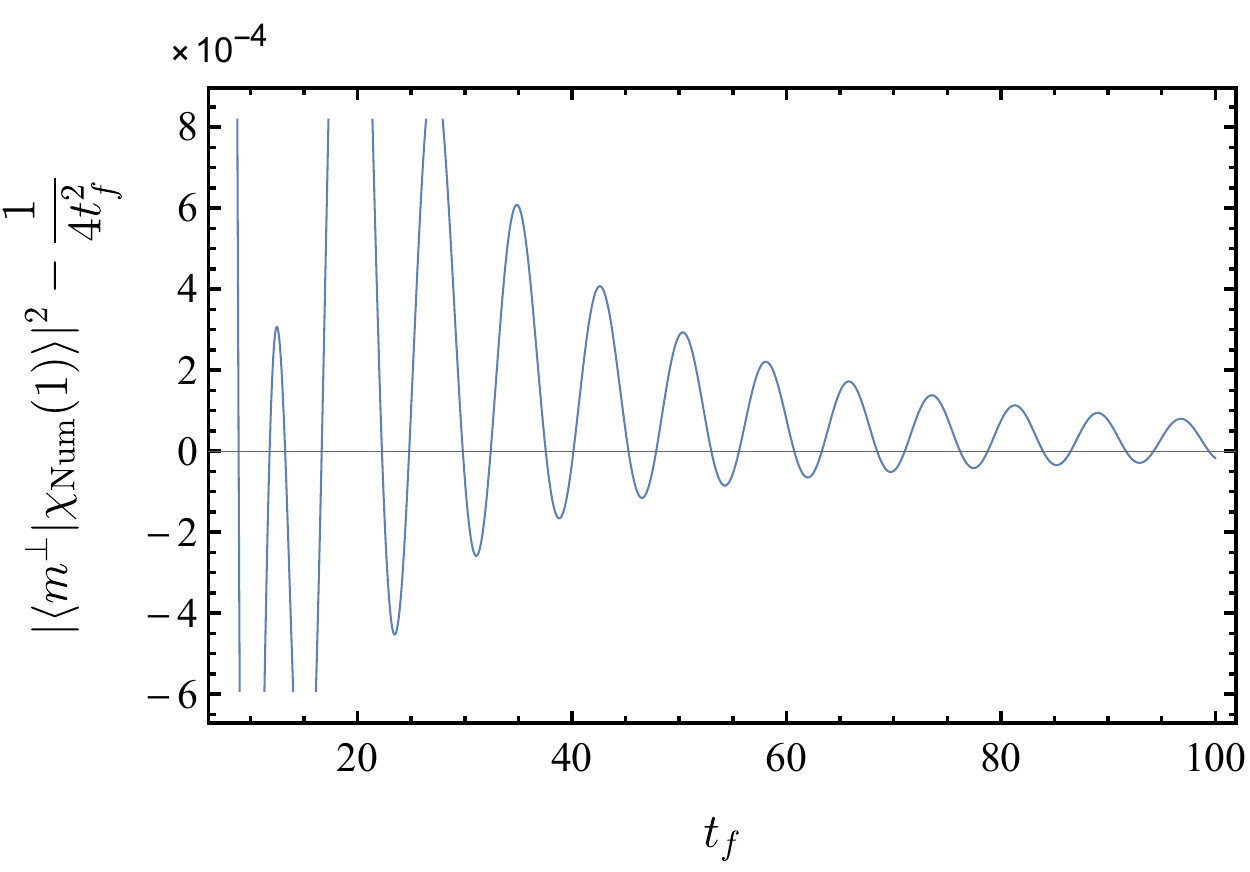}\label{fig:pgsvstf_Num_AsymptoticWKB}}
%\hspace*{\fill}
\caption{Final ground state population of a single qubit ($n=1$) in a magnetic field, under the $g_0(r)=1$ schedule. (a) Depopulation of the ground state (i.e., population in the excited state $\ket{m^\perp}$) at $r=1$ as a function of $t_f$, comparing the numerical solution ($\ket{\chi_\mathrm{Num}}$) and the two lowest orders of the WKB approximation ($\ket{\chi^{(0)}_\mathrm{WKB}}$ and $\ket{\chi^{(1)}_\mathrm{WKB}}$). The lowest order $\ket{\chi_\mathrm{WKB}^{(0)}}$ captures the asymptotic behavior of the exact solution, while $\ket{\chi_\mathrm{WKB}^{(1)}}$ becomes indistinguishable from the exact solution for $t_f \gtrsim 50$.
(b) The difference between the true population in the state $\ket{m^\perp}$ at time $r=1$ and the asymptotic prediction of $\frac{1}{4 t_f^2}$ obtained from the $1/t_f$ expansion of $\abs{\braket{m^\perp|\chi_\mathrm{WKB}^{(0)}(1)}}^2$. The asymptotic approximation becomes more accurate as $t_f$ grows.}
\label{fig:popDynamics_n=1_tf}
\end{figure*}
\begin{figure}[!htbp]
\centering
\includegraphics[width = \columnwidth]{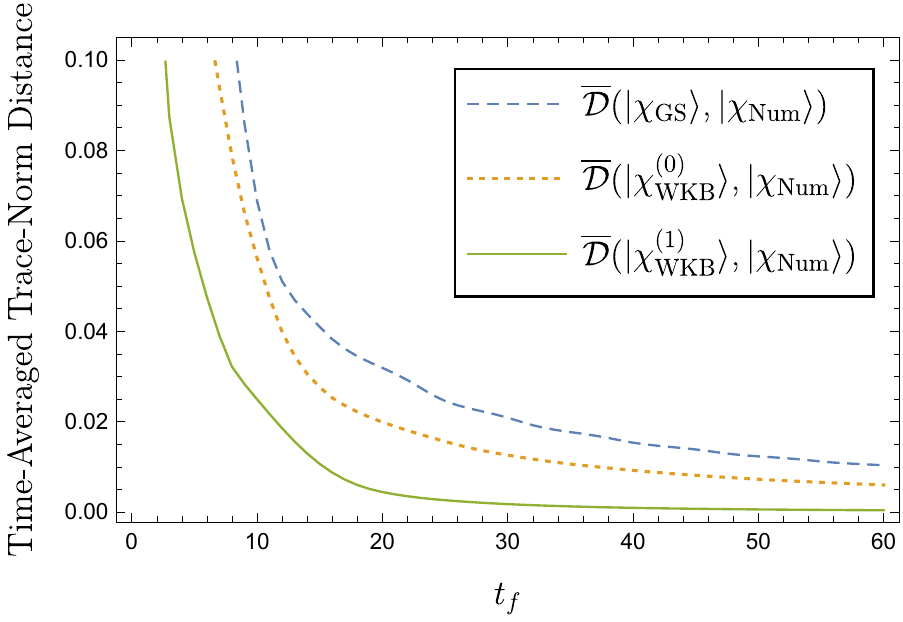}
\caption{The time-averaged trace-norm distance [see Eq.~\eqref{eqt:inttrd}] vs. $t_f$ for a single qubit ($n=1$) in a magnetic field under the $g_0(r)=1$ schedule. The distances plotted are between the numerical solution and the adiabatic approximation, and the two lowest order WKB approximations. The adiabatic and the WKB distances decrease monotonically with $t_f$, with the former being a prediction of the adiabatic theorem in the large $t_f$ limit. The WKB approximations at both orders are consistently better than the adiabatic approximation and the first-order WKB approximation improves upon the zeroth-order WKB approximation.}
%but the zeroth-order WKB approximation is worse than the adiabatic approximation for sufficiently large $t_f$, and does not decrease monotonically.
\label{fig:2ordersWKBIntError}
\end{figure}
%

%%%End of three figures.

\subsection{Single Qubit in a magnetic field} 
\label{app:qubitmagfield}

As a simple test, we first apply the formalism developed in Sec.~\ref{sec:WKBmethod} to the case $K=n=1$, which models a qubit in a time-varying magnetic field that changes from the $x$-direction to the $z$-direction, with a linear interpolation $r(s)=s$:
\beq
H(r) = - (1-r) \sigma^x - r \sigma^z,
\label{eq:1q}
\eeq
where $\sigma^x \equiv \ket{m}\bra{m^\perp} + \ket{m^\perp}\bra{m}$ and $\sigma^z \equiv \ket{m^\perp}\bra{m^\perp}-\ket{m}\bra{m}$.
Thus, the eikonal equation~\eqref{eq:23} becomes
\beq 
\label{eq:eikonalSoln1}
\theta_{\pm}^\pr = -\frac{i}{2} \left[1 \pm \Delta(r) \right],
\eeq
where $\Delta(r) \equiv \sqrt{1-2r(1-r)}$. Therefore the two energy levels of this problems are $E_\pm(s) = -i \theta_{\pm}^\pr$. Similarly, the solutions [Eqs.~\eqref{eq:28}] of the transport equations yield, after setting $K=1$, 
\bes
\begin{align} \label{eq:plainTransportSols}
\psi_0^+ &= \phi_0^- = \frac{1-r}{\sqrt{\Delta(r+\Delta)}}\\
\psi_0^- &= \phi_0^+ =\sqrt{\frac{r+\Delta}{\Delta}}\ ,
\end{align}
\ees
where we have chosen the integration constants to remove overall numerical factors.

Next, we may use these solutions to obtain the first-order correction. For this we obtain from Eq.~\eqref{eq:w'}:
\bes
\begin{align}\label{eq:plain1stSols}
\psi_1^\pm(r) &= \mp i \psi_0^\pm (r) \frac{16 r^4 - 40 r^3 +42 r^2 - 17 r + 5 \pm 6 \Delta(r)}{12(1-r) \Delta(r)^3}, \\
\phi_1^\pm(r) &= \pm i \phi_0^\pm (r) \frac{16 r^4 - 40 r^3 +42 r^2 - 17 r + 5 \mp 6 \Delta(r)}{12(1-r) \Delta(r)^3}.
\end{align}
\ees

From these expressions and the boundary conditions we construct two solutions: $\ket{\chi_\mathrm{WKB}^{(0)}}$ (using $\psi_0,\phi_0$) and $\ket{\chi_\mathrm{WKB}^{(1)}}$ (using $\psi_0,\phi_0$ and $\psi_1,\phi_1$). We expect $\ket{\chi_\mathrm{WKB}^{(1)}}$ to be a better approximation to the exact solution than $\ket{\chi_\mathrm{WKB}^{(0)}}$ and we expect the quality of approximation to improve with increasing $t_f$, i.e., with decreasing $\epsilon$. We also consider the naive adiabatic approximation, which we define as the instantaneous ground state of $H(r)$.

Figure~\ref{fig:popDynamics_n=1} shows that the WKB approximation is able to capture the correct population dynamics.\footnote{Figures were made with the help of the MaTeX package for Mathematica\textsuperscript{\textregistered} by Szabolcs Horv\'{a}t (see url: \url{http://szhorvat.net/pelican/latex-typesetting-in-mathematica.html}).} In more detail, Fig.~\ref{fig:comparetf50} shows that the approximation captures oscillations not present in a naive adiabatic approximation, and Fig.~\ref{fig:errortf50} shows that the quality of the approximation improves from the lowest order to the next order of the WKB approximation.

Next, consider the final ground state probability, $p_\mathrm{GS}(t_f)$. In Fig.~\ref{fig:2ordersWKBfinalpop}, we see that $\ket{\chi_\mathrm{WKB}^{(0)}}$ is already sufficient to capture the asymptotic scaling of $p_\mathrm{GS}$ with $t_f$. Further, $\ket{\chi_\mathrm{WKB}^{(1)}}$ captures the oscillations in $p_\mathrm{GS}(t_f)$, with an accuracy that grows with increasing $t_f$. Performing a series expansion of $\abs{\braket{1|\chi_\mathrm{WKB}^{(0)}(1)}}^2$ in powers of $\frac{1}{t_f}$, we obtain the leading order term to be $\frac{1}{4 t_f^2}$. As we see in Fig.~\ref{fig:pgsvstf_Num_AsymptoticWKB}, this asymptotic prediction is close to the asymptotic scaling of the numerical solution.

Finally, consider the time-averaged trace-norm distance between two time-evolving states $\ket{\chi_1(t)}$ and $\ket{\chi_2(t)}$:
\bes
\label{eqt:inttrd}
\begin{align}
\overline{\mathcal{D}}\left(\ket{\chi_1},\ket{\chi_2}\right) &= \frac{1}{t_f}\int_0^{t_f} \mathcal{D}\left(\ket{\chi_1},\ket{\chi_2}\right) dt\\
\mathcal{D}\left(\ket{\chi_1},\ket{\chi_2}\right) & \equiv \frac{1}{2} \left\|\ket{\chi_1}\bra{\chi_1}-\ket{\chi_2}\bra{\chi_2}\right\|_1 \ .
\end{align}
 \ees
The results comparing the numerical solution to the naive adiabatic approximation and the two lowest WKB approximation orders are shown in Fig.~\ref{fig:2ordersWKBIntError}. As expected, the naive adiabatic approximation becomes better as $t_f$ increases, and the same is true for the WKB approximations, which are both more accurate than the adiabatic approximation. Moreover, the first-order WKB approximation is better than the adiabatic approximation according to the time-averaged trace-norm distance metric. 

%We revisit this point below and show that in fact the situation improves with increasing $n$.

Taken together, the results for the $n=1$ case show that both the zeroth-order WKB approximation and the first-order WKB approximation consistently improve upon the naive adiabatic approximation, and the first-order WKB approximation can be used to pick out more subtle features of the quantum evolution.

%%%%Three figures below
%

%
\begin{figure}[!h]
{\includegraphics[width = \columnwidth]{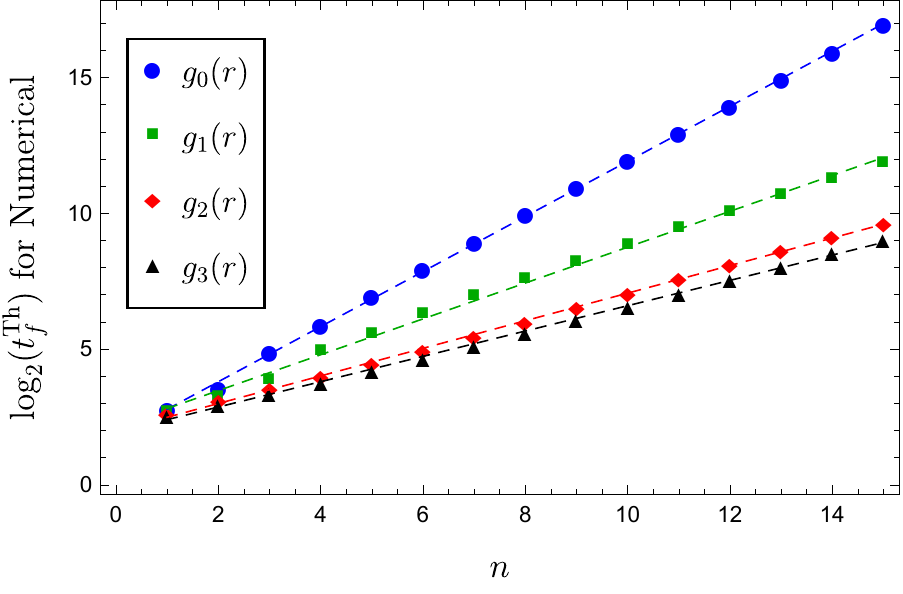}}
\caption{The time required to achieve a final ground state probability of $0.95$ for the schedules defined in Eqs.~\eqref{eq:4schedsprime} ($\log$ scale). The straight lines represent exponential scaling fits of $\mathcal{O}(2^{1.01n})$,  $\mathcal{O}(2^{0.667n})$, $\mathcal{O}(2^{0.508n})$, and $\mathcal{O}(2^{0.463n})$ respectively.}
\label{fig:manyschedsSchro}
\end{figure}

\begin{figure}[!h]
\centering
\includegraphics[width = \columnwidth]{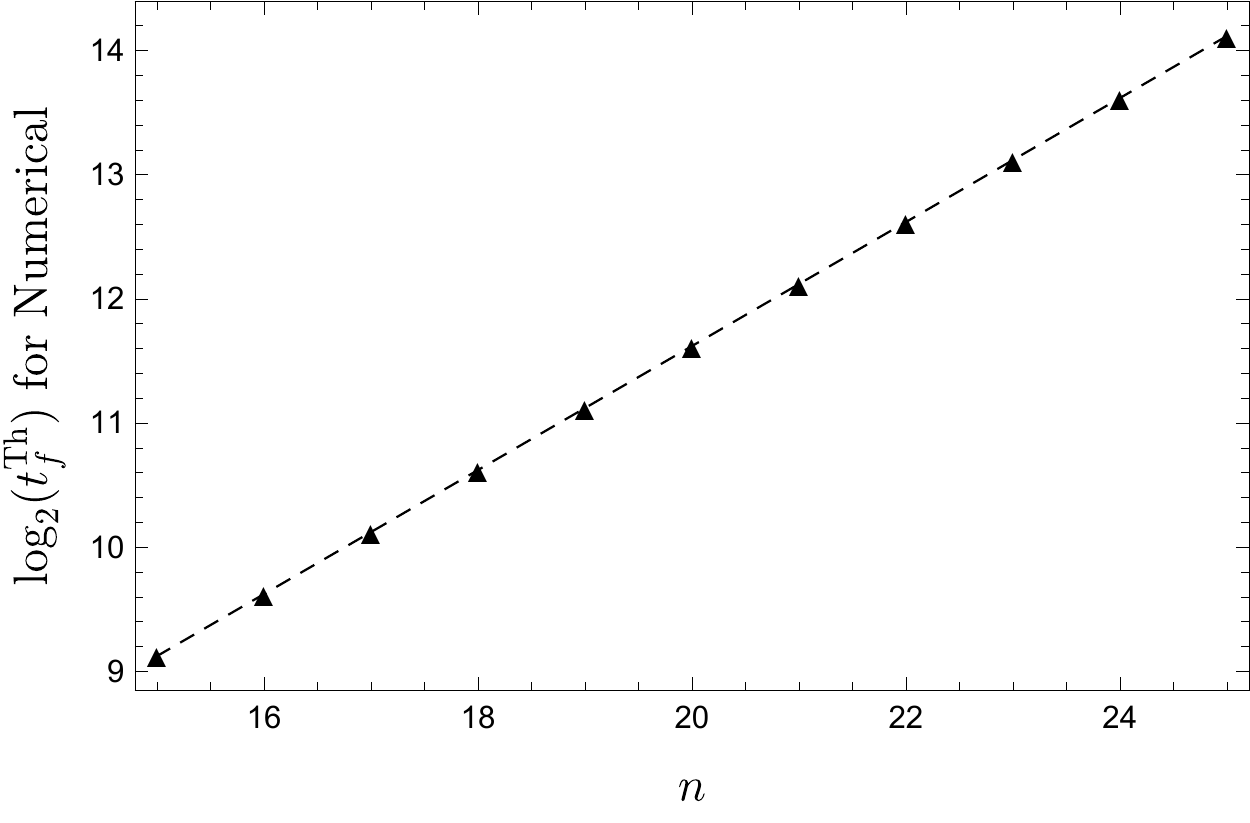}
\caption{The scaling of $p_\mathrm{GS}(t_f)$ from the numerically exact solution under the $g_3$ schedule, for larger problem sizes than in Fig.~\ref{fig:manyschedsSchro}. The straight line represents an exponential scaling fit of $\mathcal{O}(2^{0.499n})$. Thus, the scaling converges to the expected scaling of $\mathcal{O}(2^{n/2})$ predicted by the query complexity bound  \cite{Bennett:1997lh}.}
\label{fig:TimeToThresholdg3}
\end{figure}

\begin{figure}[!h]
{\includegraphics[width = \columnwidth]{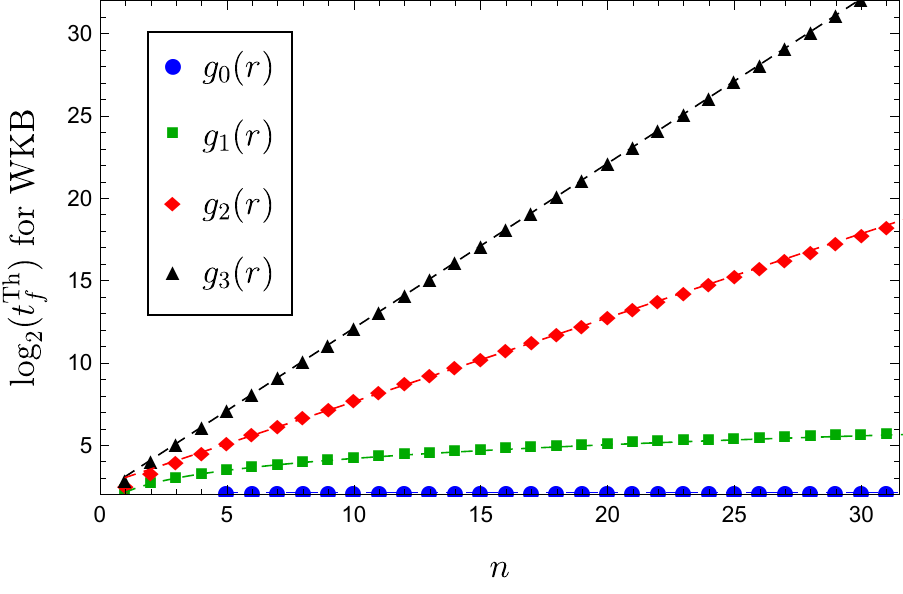}}
\caption{The time required to achieve a final ground state probability of $0.95$ for the interpolations defined in Eqs.~\eqref{eq:4schedsprime}] ($\log$ scale), using the WKB approximation at the lowest order. The straight lines represent fits of $\mathcal{O}(2^{n})$,  $\mathcal{O}(2^{0.51n})$, $\mathcal{O}(2^{3.5n^{0.2}})$, and $\mathcal{O}(1)$ respectively. Thus, the lowest-order WKB approximation predicts the right scaling only for the optimized schedule $g_2(r)$.}
\label{fig:manyschedsWKB}
\end{figure}

\begin{figure*}[t]
%\hspace*{\fill}
\subfigure[\ ]{\includegraphics[width = \columnwidth]{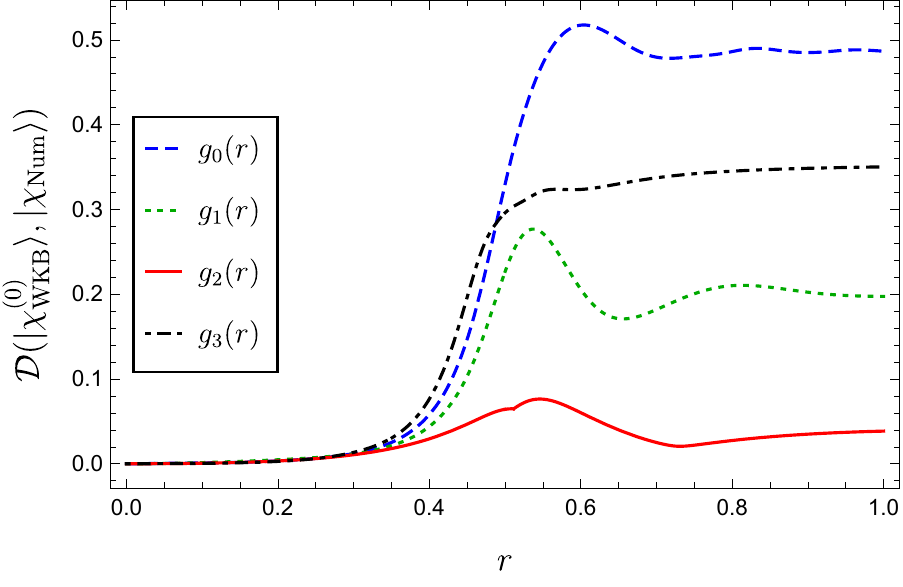}\label{fig:manyschedsTrDvsrWKBNum}}\hfill
\subfigure[\ ]{\includegraphics[width = \columnwidth]{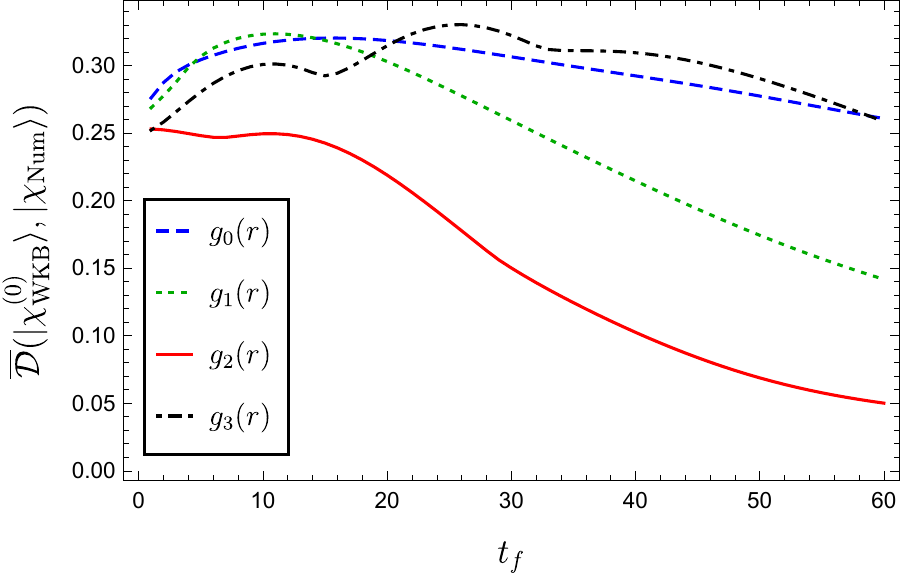}\label{fig:manyschedsITDWKBNum}}
%\hspace*{\fill}
\caption{(a) The trace-norm distance between the lowest order WKB approximation and the numerically exact solution for the four different schedules [Eq.~\eqref{eq:4schedsprime}] as a function of the evolution parameter $r$. Here $n=6$ and $t_f=60$. (b) The time-averaged trace-norm distance [Eq.~\eqref{eqt:inttrd}] between the lowest order WKB approximation and the numerically exact solution for the four different schedules [Eqs.~\eqref{eq:4schedsprime}]. Here $n=6$. Both panels are consistent with Fig.~\ref{fig:manyschedsWKB} where $g_2$ recovers the correct scaling. Recall that $g_2$ represents the optimal schedule found in Ref.~\cite{Roland:2002ul}, which provides the best approximation to the numerical evolution.}
\label{fig:manyschedsTrDvsrWKBNum-full}
\end{figure*}

\begin{figure*}[!htbp]%Pgs vs t_f
\hspace*{\fill}
\subfigure{\includegraphics[width = \columnwidth]{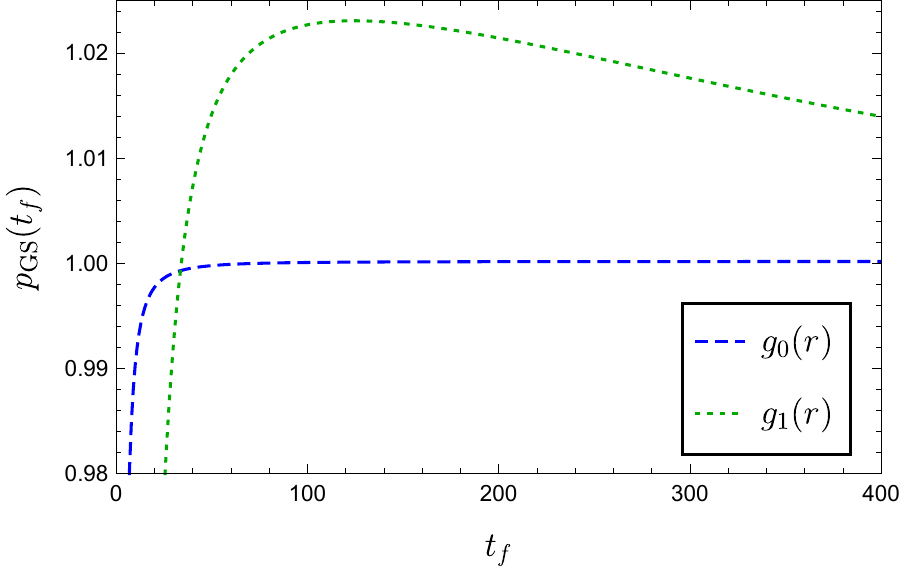}\label{fig:pgsvstfg0g1}} \hfill
\subfigure{\includegraphics[width = \columnwidth]{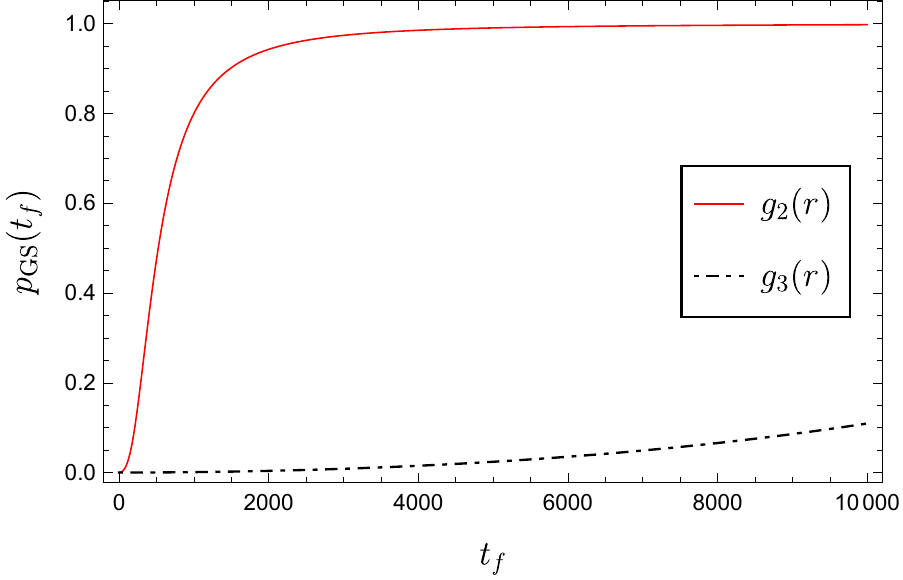}\label{fig:pgsvstfg2g3}}
\hspace*{\fill}
\caption{Final ground state probability $p_\mathrm{GS}$ as function of total evolution time $t_f$ for the Grover problem with $n=4$ for the four different schedules, $g_\alpha$ with $\alpha \in \{0,1,2,3\}$ as predicted by the WKB approximation at first order. (a) The $g_0$ and $g_1$ schedules. The rise in $p_\mathrm{GS}$ as a function of $t_f$ is very steep, and quickly exceeds $1$ for both schedules (the $g_0$ curve goes very slightly above $1$). The $g_0$ curve rises faster than the $g_1$ curve for $p_\mathrm{GS}\leq 1$.
(b) The $g_2$ and $g_3$ schedules.The rise in the $p_\mathrm{GS}$ curve for $g_2$ is much steeper than the rise for the $g_3$ curve. In general, the smaller is $\alpha$, the larger the steepness in the $p_\mathrm{GS}(t_f)$ curve. This is consistent with the $t_f^\mathrm{Th}(n)$ scalings obtained in Fig.~\ref{fig:manyschedsWKB}.}
\label{fig:pgsvstf}
\end{figure*}

\begin{figure}[!htbp]
\centering
\includegraphics[width = \columnwidth]{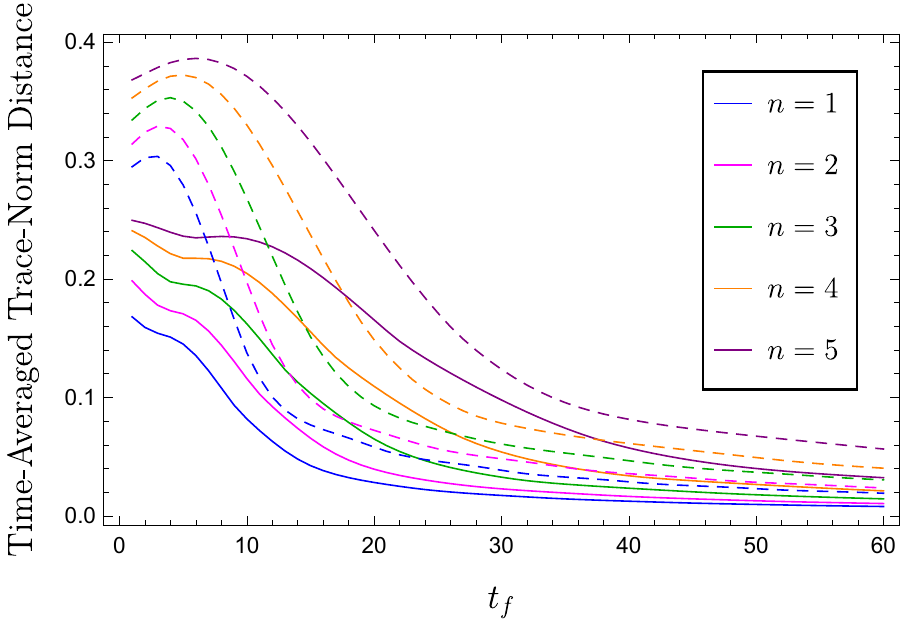}\hfill
\caption{The Grover problem with the $g_2(r)$ schedule for $n \in \{1,\dots, 5\}$. (a) The time-averaged trace-norm distance [Eq.~\eqref{eqt:inttrd}] between the numerical solution and the lowest order WKB approximation (solid), i.e, $\overline{\mathcal{D}}(\ket{\chi_\mathrm{WKB}^{(0)}},\ket{\chi_\mathrm{Num}})$; and the time-averaged trace-distance between the numerical solution and the adiabatic approximation (dashed), i.e, $\overline{\mathcal{D}}(\ket{\chi_\mathrm{GS}},\ket{\chi_\mathrm{Num}})$. 
%Notice that the solid lines of a particular color are always below (and hence a better approximation) than the dashed lines of the same color. Thus, 
The lowest order WKB approximation is always better than the adiabatic approximation for all $n$ values we have tested. (For both the solid and dashed lines, the lower the line on the plot, the lower the value of $n$.)}
\label{fig:inttracedistNumWKBAdia}
\end{figure}

%%
%\begin{figure*}[!htbp]%Crossover timescales
%\hspace*{\fill}
%\subfigure[\ ]{\includegraphics[width = 0.9\columnwidth]{n=1to5_g2_WKBNum_NumAdia_IntTrD}\label{fig:inttracedistNumWKBAdia}}\hfill
%\subfigure[\ ]{\includegraphics[width = 0.9\columnwidth]{tfcrssvr_vs_n_g2_WKB0}\label{fig:tfcrssvrWKBAdia}}
%\hspace*{\fill}
%\caption{The Grover problem with the $g_2(r)$ schedule for $n \in \{1,\dots, 5\}$. (a) The time-averaged trace-norm distance [Eq.~\eqref{eqt:inttrd}] between the numerical solution and the lowest order WKB approximation (solid), i.e, $\mathcal{D}(\ket{\chi_\mathrm{WKB}^{(0)}},\ket{\chi_\mathrm{Num}})$; and the time-averaged trace-distance between the numerical solution and the adiabatic approximation (dashed), i.e, $\mathcal{D}(\ket{\chi_\mathrm{GS}},\ket{\chi_\mathrm{Num}})$. (b) The crossover time $t_f^\mathrm{crssvr}$ at which the WKB approximation becomes a poorer approximation to the true evolution than the adiabatic approximation, as a function of the number of qubits $n$, and extracted from the crossings observed in (a). The time-scale for which the WKB approximation is better than the adiabatic approximation increases with problem size. The scaling of $t_f^\mathrm{crssvr}(n)$ is $\mathcal{O}(2^{0.34 n})$.}
%\label{fig:crossovertimescales}
%\end{figure*}

%%%%End of three figures.

\subsection{The $n$-qubit Grover problem} 
\label{sec:n-Grover}

We next turn to a study of the Grover problem as a function of problem size $n$, with $n>1$. The quantity of interest to us is how long we need to run the adiabatic algorithm before a certain threshold probability of success $p_\mathrm{Th}$ is exceeded. The associated threshold timescale is defined as:
\beq \label{eq:tfThdef}
t_f^\mathrm{Th} \equiv \min \{t_f : p_\mathrm{GS}(t) > p_\mathrm{Th} \ \forall  t>t_f \}.
\eeq
Here, $p_\mathrm{GS}(t)$ represents the probability of finding the ground state at the end an adiabatic evolution of time $t$. We choose $p_\mathrm{Th}=0.95$ (we have checked that the results are insensitive to changing $p_\mathrm{Th}$).

First, in Fig.~\ref{fig:manyschedsSchro} we show how $t_f^\mathrm{Th}(n)$ scales for the numerical solution, under the four different schedules defined in Eqs.~\eqref{eq:4schedsprime}. 
It appears as though the scaling for the $g_3(r)$ schedule is better than the theoretically optimal scaling of $2^{n/2}$ \cite{Bennett:1997lh}, but this is a small $n$ effect as shown in Fig.~\ref{fig:TimeToThresholdg3}. 

Next, we examine how well the WKB approximation does in predicting these scalings. In Fig.~\ref{fig:manyschedsWKB} we plot the scaling of $t_f^\mathrm{Th}(n)$ for the same four schedules, under the lowest order WKB approximation. Only the $g_2(r)$ schedule (which slows as the inverse-square of the gap) yields the correct scaling of $t_f^\mathrm{Th}(n)$. This is also the schedule which yields the smallest instantaneous and time-averaged trace-norm distance, as shown in Figs.~\ref{fig:manyschedsTrDvsrWKBNum} and \ref{fig:manyschedsITDWKBNum}, respectively. For the other schedules, Fig.~\ref{fig:manyschedsWKB} shows that the WKB approximation gives answers that are dramatically different from the exact solution. Furthermore, for the $g_0$ and $g_1$ schedules, the scaling with $n$ of $t_f^\mathrm{Th}(n)$ violates the query complexity bound \cite{Bennett:1997lh}. 

Why do the approximations for the $g_0$, $g_1$, and $g_3$ schedules give us the wrong scalings, while the approximation for the $g_2$ schedule gives us the correct scaling? A possible answer lies in the steepness of the final-time approximate success probability curves for the different schedules. In Fig.~\ref{fig:pgsvstf} we show the $p_\mathrm{GS}(t_f)$ curves for all four schedules for $n=4$ ($K=15$) as predicted by the first-order WKB approximation (the highest order at which we are able to obtain analytic expressions). For the $g_0$ and $g_1$ schedules, we see that the final ground state probability rises very sharply and exceeds unity (very slightly for $g_0$), and thereby becomes nonsensical [see Fig.~\ref{fig:pgsvstfg0g1}], while for the $g_2$ and $g_3$ schedules,  $p_\mathrm{GS}(t_f)\leq 1$ [see Fig.~\ref{fig:pgsvstfg2g3}]. Further, we observe that the curves are ordered from steepest to shallowest rise as $g_0$, $g_1$, $g_2$, $g_3$. We conjecture that this rise in $p_\mathrm{GS}$ with $t_f$ continues to slow down with increasing $\alpha$. Thus the $g_2$ schedule captures the right scaling [in Fig.~\ref{fig:manyschedsWKB}] by capturing the right steepness: for $\alpha <2$ the rise is too steep, and for $\alpha > 2$ the rise is too shallow. A full explanation of this phenomenon is left to future work, but we speculate that the $g_0$ and $g_1$ schedules correspond to effective Hamiltonians that no longer represent the Grover problem.

Given that the WKB approximation gives consistent results only for the $g_2(r)$ schedule, we focus on this schedule and examine where the WKB approximation performs better than the naive adiabatic approximation. As can be seen in Fig.~\ref{fig:inttracedistNumWKBAdia}, for the $g_2(r)$ schedule, the WKB approximation always has an advantage over the naive adiabatic approximation. Further, it is clear that the advantage is bigger for smaller evolution times $t_f$ and for larger problem sizes $n$.

As we have indicated above, the WKB approximants are generically not normalized: they can be sub-normalized or super-normalized. Two questions arise: (1) Is the degree of non-normalization a good indicator of approximation quality of the WKB aproximation? (2) Does renormalization by fiat improve the quality of the approximation? 
%We address these questions next. 

%\subsubsection{Is non-normalization a good indicator of approximation quality?}\label{sec:renorm}

\begin{figure}[!htbp]
\centering
\includegraphics[width = \columnwidth]{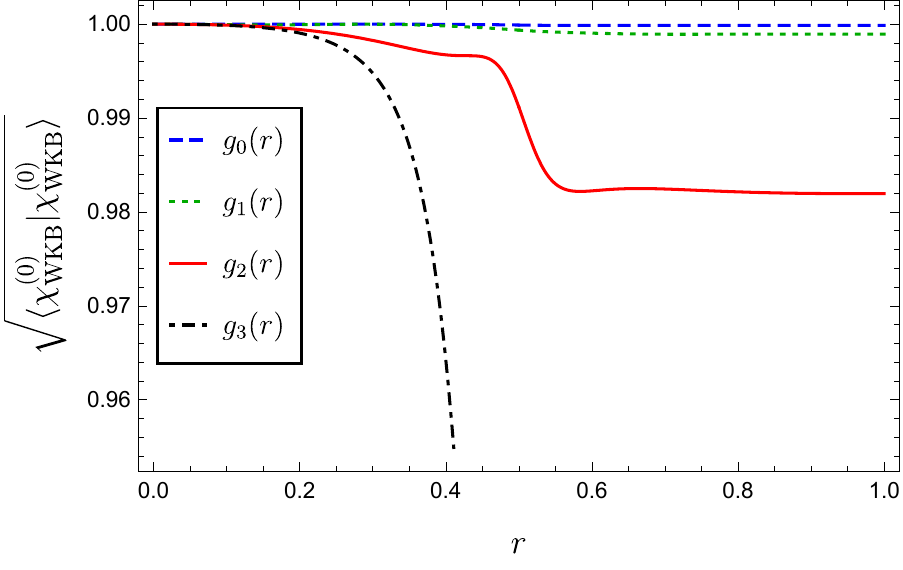}
\caption{The norm of the WKB approximation at the lowest order as a function of the evolution parameter $r$ for different schedules. $g_2$ represents the optimal schedule, but does not maintain normalization. Note that $g_3$ is significantly more sub-normalized that than the rest, and dips down to about $0.7$ (not shown). Here $n=6$ and $t_f=60$.}
\label{fig:manyschedsNormvsr}
\end{figure}

First, in Fig.~\ref{fig:manyschedsNormvsr} we plot the norm of the WKB approximation at the lowest order for the case of $n=6$ and $t_f=60$ as a function of the anneal parameter $r$ for all four schedules. In this regime, the WKB approximation is sub-normalized for all schedules. Arranging the schedules from farthest from normalization to closest to normalization, we have: $g_3,g_2,g_1,g_0$, with $g_0$ and $g_1$ closest to being normalized.
%(Note that $g_3$ is significantly more sub-normalized that than the rest: see Fig.~\ref{fig:g3Normvsr}.)
%
On the other hand, we have seen [Fig.~\ref{fig:manyschedsTrDvsrWKBNum} and Figs.~\ref{fig:manyschedsWKB},\ref{fig:manyschedsITDWKBNum}] that the best approximation was obtained for the $g_2$ schedule. 
%But the $g_2$ schedule is not the schedule which is closest to being normalized. Indeed, both $g_0$ and $g_1$ are closer to being normalized than $g_2$. 
Thus, we conclude that the degree of non-normalization is not a good indicator of the quality of approximation.

%\begin{figure}[!htbp]
%\centering
%\includegraphics[width = \columnwidth]{Normvsr_tf=60_k=63_g3schedule}
%\caption{The norm of the WKB approximation at the lowest order as a function of the evolution parameter $r$ for the $g_3(r)$ schedule. The WKB approximant for this schedule is the most sub-normalized one among the four schedules we've studied. Here $n=6$ and $t_f=60$.}
%\label{fig:g3Normvsr}
%\end{figure}

%\subsubsection{Does renormalization improve approximation quality?}

Second, we consider renormalization of the WKB approximation, by which we mean:
\beq
\ket{\chi_\mathrm{rWKB}} \equiv \frac{\ket{\chi_\mathrm{WKB}}}{\sqrt{\braket{\chi_\mathrm{WKB}|\chi_\mathrm{WKB}}}}\ ,
\eeq
where we have denoted the renormalized WKB approximants as $\ket{\chi_\mathrm{rWKB}}$.
Note that it is somewhat \emph{ad hoc} to normalize the approximant: a renormalization step is not part of the standard WKB approximation procedure. With this caveat, we now analyze the behavior of the WKB approximants after renormalization. 
 
\begin{figure}[t]
\centering
\includegraphics[width = \columnwidth]{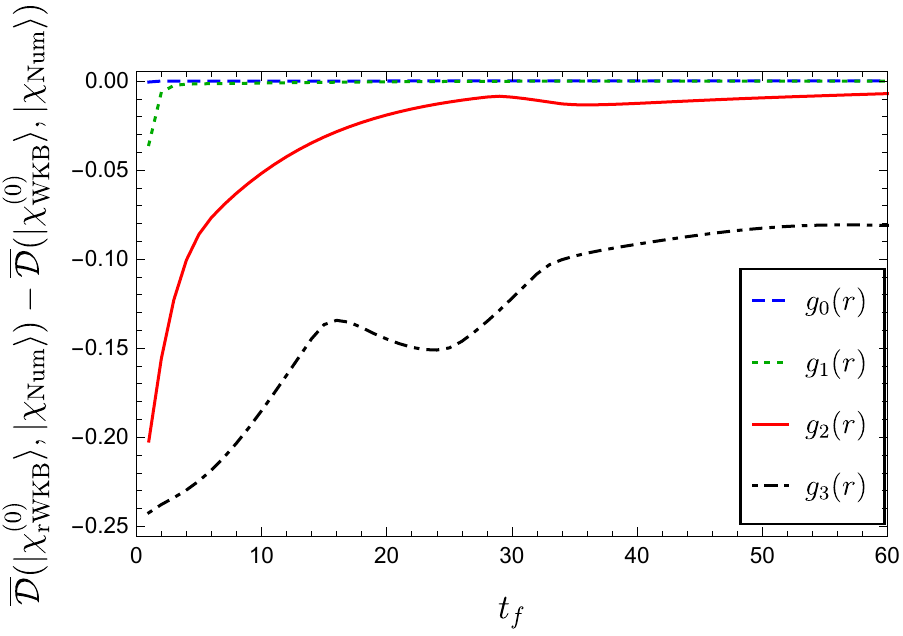}
\caption{The difference between the time-averaged trace-norm distances for the renormalized and unnormalized WKB approximants, for the four different schedules [Eqs.~\eqref{eq:4schedsprime}].  Here $n=6$ and $t_f=60$. A negative value means that the unnormalized WKB approximation deviates more from the numerically exact solution than the renormalized WKB approximation.}
\label{fig:manyschedsITDdiffs}
\end{figure}

First we consider the time-averaged trace-norm distance, $\overline{\mathcal{D}}$ [Eq.~\eqref{eqt:inttrd}]. As shown in Fig.~\ref{fig:manyschedsITDdiffs}, renormalizing the WKB approximation does improve the approximation. 
%But unfortunately, as we saw with regards to $t_f^\mathrm{Th}(n)$, this does not directly translate into improved predictions for $t_f^\mathrm{Th}(n)$. That is, the improvement in $\mathcal{D}$ translated into the correct scaling for $t_f^\mathrm{Th}(n)$ only for the $g_3$ schedule; whereas for the $g_0,g_1,$ and $g_2$ schedules, the improvement in $\mathcal{D}$ translated into degradation of the predictions for $t_f^\mathrm{Th}(n)$.

\begin{figure}[t]
{\includegraphics[width = \columnwidth]{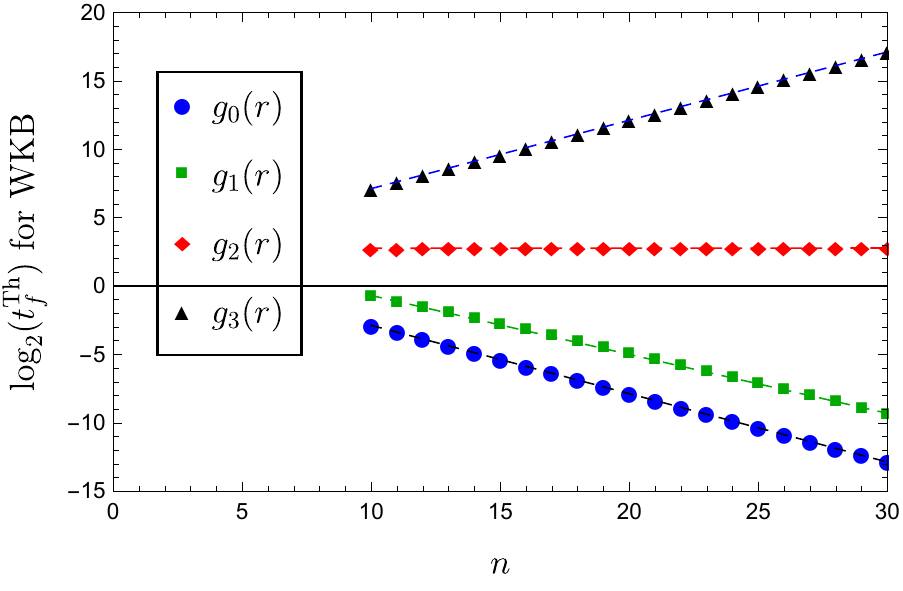}}
\caption{The time required to achieve a final ground state probability of $0.95$ for the interpolations defined in Eqs.~\eqref{eq:4schedsprime}] ($\log$ scale), using the renormalized version WKB approximation at the lowest order. The straight lines represent fits of $\mathcal{O}(2^{n/2})$,  $\mathcal{O}(1)$, $\mathcal{O}(2^{-n/2})$, and $\mathcal{O}(2^{-n/2})$ respectively.}
\label{fig:manyschedsrenormedWKB}
\end{figure}

However, the situation changes when we consider the threshold timescale $t_f^\mathrm{Th}(n)$ [Eq.~\eqref{eq:tfThdef}], shown in Fig.~\ref{fig:manyschedsrenormedWKB}. We see that renormalizing the WKB approximation yields highly unphysical results for the $g_0,g_1$, and $g_2$ schedules. In particular, for the $g_0$ and $g_1$ schedules, we see that the renormalized WKB approximation predicts a scaling for $t_f^\mathrm{Th}(n)$ that \emph{decreases} with problem size $n$. For the $g_2$ schedule we see a scaling of $\mathcal{O}(1)$. So, while the unnormalized WKB predicted the correct scaling for the $g_2$ schedule, the renormalized WKB does not retain that feature. On the other hand, for the $g_3$ schedule, we see that the renormalized WKB predicts the correct scaling of $\mathcal{O}(2^{n/2})$, fixing the incorrect scaling of the unnormalized WKB approximation for that schedule.

\section{Summary and Conclusions}
\label{sec:conc}

We have presented a straightforward technique to obtain an analytic asymptotic approximation to slowly evolving $2$-level systems by adapting the WKB method. We have applied it to a problem that is motivated by adiabatic quantum computation: the Hamiltonian Grover search problem. This problem has a physical Hilbert space of dimension $2^n$, but is effectively constrained to a $2$-dimensional subspace. We have seen that in this case when $n=1$, the WKB method provides good approximations, especially to the population dynamics. We saw that the WKB approximation can capture fluctuations in the population that are absent in the purely adiabatic (ground state) evolution. Thus, the WKB is quasi-adiabatic. For completeness, in Appendix~\ref{app:Hage}, we compare our WKB approximation to the asymptotic expansion method of Hagedorn and Joye~\cite{Hagedorn:2002kx}, and show that the latter misses the oscillations that are captured by the quasi-adiabatic WKB expansion.

Turning to the Grover problem with $n>1$ and with different interpolation schedules, we observed that the WKB approximation yields meaningful results only for the schedule which slows quadratically with the ground state gap. For this $g_2(r)$ schedule, the WKB approximation is able to capture the scaling with $n$ of $t_f^\mathrm{Th}$, and hence recovers the quantum speedup, even at the lowest approximation order. On the other hand, for the schedules that slow down more slowly than quadratically with the gap, the WKB approximation violates normalization and predicts an impossible faster-than-quadratic quantum speedup for the Grover problem. 

We also saw that, using the time-averaged trace-norm distance, for the $g_2(r)$ schedule, the WKB approximation always does better than a naive adiabatic approximation, and the advantage becomes more pronounced for larger system sizes and shorter evolution times. 

Turning to the question of the whether the norm of the WKB approximation is a good signal of approximation quality, we saw that this is not the case. Further, we saw that enforcing renormalization by fiat gives mixed results. On the one hand, it lead to an improvement in the time-averaged trace-distance and also gave the right predictions for scaling of the threshold timescale for the $g_3$ schedule. On the other hand, it gave incorrect predictions for the scaling of the threshold timescales for the $g_0,g_1,$ and $g_2$ schedules, especially degrading the prediction for the $g_2$ schedule compared to its unnormalized counterpart.

An interesting problem for future work is to provide more rigorous justifications and explanations for when and where the WKB approximation provides good approximations. With a better understanding of the regions where the WKB approximation performs well, and if the approximation errors are better controlled, the method could be used in the design of quantum control protocols to implement quantum gates~\cite{Peirce:88,d2001optimal,Grace:2007aa,Barnes:2012fk,palittapongarnpim2017learning}.

\acknowledgements
We are grateful to an anonymous referee who provided several constructive suggestions. We also thank Tameem Albash and Milad Marvian for helpful discussions and comments. This work was supported under ARO grant number W911NF-12-1-0523 and NSF grant number INSPIRE-1551064.

\bibliographystyle{apsrev4-1}
\bibliography{refs}

\begin{thebibliography}{28}%
\makeatletter
\providecommand \@ifxundefined [1]{%
 \@ifx{#1\undefined}
}%
\providecommand \@ifnum [1]{%
 \ifnum #1\expandafter \@firstoftwo
 \else \expandafter \@secondoftwo
 \fi
}%
\providecommand \@ifx [1]{%
 \ifx #1\expandafter \@firstoftwo
 \else \expandafter \@secondoftwo
 \fi
}%
\providecommand \natexlab [1]{#1}%
\providecommand \enquote  [1]{``#1''}%
\providecommand \bibnamefont  [1]{#1}%
\providecommand \bibfnamefont [1]{#1}%
\providecommand \citenamefont [1]{#1}%
\providecommand \href@noop [0]{\@secondoftwo}%
\providecommand \href [0]{\begingroup \@sanitize@url \@href}%
\providecommand \@href[1]{\@@startlink{#1}\@@href}%
\providecommand \@@href[1]{\endgroup#1\@@endlink}%
\providecommand \@sanitize@url [0]{\catcode `\\12\catcode `\$12\catcode
  `\&12\catcode `\#12\catcode `\^12\catcode `\_12\catcode `\%12\relax}%
\providecommand \@@startlink[1]{}%
\providecommand \@@endlink[0]{}%
\providecommand \url  [0]{\begingroup\@sanitize@url \@url }%
\providecommand \@url [1]{\endgroup\@href {#1}{\urlprefix }}%
\providecommand \urlprefix  [0]{URL }%
\providecommand \Eprint [0]{\href }%
\providecommand \doibase [0]{http://dx.doi.org/}%
\providecommand \selectlanguage [0]{\@gobble}%
\providecommand \bibinfo  [0]{\@secondoftwo}%
\providecommand \bibfield  [0]{\@secondoftwo}%
\providecommand \translation [1]{[#1]}%
\providecommand \BibitemOpen [0]{}%
\providecommand \bibitemStop [0]{}%
\providecommand \bibitemNoStop [0]{.\EOS\space}%
\providecommand \EOS [0]{\spacefactor3000\relax}%
\providecommand \BibitemShut  [1]{\csname bibitem#1\endcsname}%
\let\auto@bib@innerbib\@empty
%</preamble>
\bibitem [{\citenamefont {Farhi}\ and\ \citenamefont
  {Gutmann}(1998)}]{FarhiAnalog}%
  \BibitemOpen
  \bibfield  {author} {\bibinfo {author} {\bibfnamefont {E.}~\bibnamefont
  {Farhi}}\ and\ \bibinfo {author} {\bibfnamefont {S.}~\bibnamefont
  {Gutmann}},\ }\href {\doibase 10.1103/PhysRevA.57.2403} {\bibfield  {journal}
  {\bibinfo  {journal} {Phys. Rev. A}\ }\textbf {\bibinfo {volume} {57}},\
  \bibinfo {pages} {2403} (\bibinfo {year} {1998})}\BibitemShut {NoStop}%
\bibitem [{\citenamefont {Farhi}\ \emph {et~al.}(2000)\citenamefont {Farhi},
  \citenamefont {Goldstone}, \citenamefont {Gutmann},\ and\ \citenamefont
  {Sipser}}]{Farhi:00}%
  \BibitemOpen
  \bibfield  {author} {\bibinfo {author} {\bibfnamefont {E.}~\bibnamefont
  {Farhi}}, \bibinfo {author} {\bibfnamefont {J.}~\bibnamefont {Goldstone}},
  \bibinfo {author} {\bibfnamefont {S.}~\bibnamefont {Gutmann}}, \ and\
  \bibinfo {author} {\bibfnamefont {M.}~\bibnamefont {Sipser}},\ }\href
  {http://arxiv.org/abs/quant-ph/0001106} {\bibfield  {journal} {\bibinfo
  {journal} {arXiv:quant-ph/0001106}\ } (\bibinfo {year} {2000})}\BibitemShut
  {NoStop}%
\bibitem [{\citenamefont {Feynman}(1985)}]{Feynman:1985ul}%
  \BibitemOpen
  \bibfield  {author} {\bibinfo {author} {\bibfnamefont {R.~P.}\ \bibnamefont
  {Feynman}},\ }\bibfield  {booktitle} {\emph {\bibinfo {booktitle} {Optics
  News}},\ }\href {\doibase 10.1364/ON.11.2.000011} {\bibfield  {journal}
  {\bibinfo  {journal} {Optics News}\ }\textbf {\bibinfo {volume} {11}},\
  \bibinfo {pages} {11} (\bibinfo {year} {1985})}\BibitemShut {NoStop}%
\bibitem [{\citenamefont {Peres}(1985)}]{Peres:1985aa}%
  \BibitemOpen
  \bibfield  {author} {\bibinfo {author} {\bibfnamefont {A.}~\bibnamefont
  {Peres}},\ }\href {http://link.aps.org/doi/10.1103/PhysRevA.32.3266}
  {\bibfield  {journal} {\bibinfo  {journal} {Physical Review A}\ }\textbf
  {\bibinfo {volume} {32}},\ \bibinfo {pages} {3266} (\bibinfo {year}
  {1985})}\BibitemShut {NoStop}%
\bibitem [{\citenamefont {Margolus}(1990)}]{Margolus:90}%
  \BibitemOpen
  \bibfield  {author} {\bibinfo {author} {\bibfnamefont {N.}~\bibnamefont
  {Margolus}},\ }in\ \href {http://people.csail.mit.edu/nhm/pqc.pdf} {\emph
  {\bibinfo {booktitle} {Complexity, Entropy, and the Physics of
  Information}}},\ \bibinfo {series} {SFI Studies in the Sciences of
  Complexity}, Vol.\ \bibinfo {volume} {VIII},\ \bibinfo {editor} {edited by\
  \bibinfo {editor} {\bibfnamefont {W.}~\bibnamefont {Zurek}}}\ (\bibinfo
  {publisher} {Addison-Wesley},\ \bibinfo {year} {1990})\ pp.\ \bibinfo {pages}
  {273--287}\BibitemShut {NoStop}%
\bibitem [{\citenamefont {Albash}\ and\ \citenamefont
  {Lidar}(2016)}]{Albash-Lidar:RMP}%
  \BibitemOpen
  \bibfield  {author} {\bibinfo {author} {\bibfnamefont {T.}~\bibnamefont
  {Albash}}\ and\ \bibinfo {author} {\bibfnamefont {D.~A.}\ \bibnamefont
  {Lidar}},\ }\href {http://arXiv.org/abs/1611.04471} {\bibfield  {journal}
  {\bibinfo  {journal} {arXiv:1611.04471}\ } (\bibinfo {year}
  {2016})}\BibitemShut {NoStop}%
\bibitem [{\citenamefont {Roland}\ and\ \citenamefont
  {Cerf}(2002)}]{Roland:2002ul}%
  \BibitemOpen
  \bibfield  {author} {\bibinfo {author} {\bibfnamefont {J.}~\bibnamefont
  {Roland}}\ and\ \bibinfo {author} {\bibfnamefont {N.~J.}\ \bibnamefont
  {Cerf}},\ }\href {http://link.aps.org/doi/10.1103/PhysRevA.65.042308}
  {\bibfield  {journal} {\bibinfo  {journal} {Phys. Rev. A}\ }\textbf {\bibinfo
  {volume} {65}},\ \bibinfo {pages} {042308} (\bibinfo {year}
  {2002})}\BibitemShut {NoStop}%
\bibitem [{\citenamefont {Jansen}\ \emph {et~al.}(2007)\citenamefont {Jansen},
  \citenamefont {Ruskai},\ and\ \citenamefont {Seiler}}]{Jansen:07}%
  \BibitemOpen
  \bibfield  {author} {\bibinfo {author} {\bibfnamefont {S.}~\bibnamefont
  {Jansen}}, \bibinfo {author} {\bibfnamefont {M.-B.}\ \bibnamefont {Ruskai}},
  \ and\ \bibinfo {author} {\bibfnamefont {R.}~\bibnamefont {Seiler}},\ }\href
  {http://scitation.aip.org/content/aip/journal/jmp/48/10/10.1063/1.2798382}
  {\bibfield  {journal} {\bibinfo  {journal} {J. Math. Phys.}\ }\textbf
  {\bibinfo {volume} {48}},\ \bibinfo {pages} {102111} (\bibinfo {year}
  {2007})}\BibitemShut {NoStop}%
\bibitem [{\citenamefont {Grover}(1996)}]{Grover:1996}%
  \BibitemOpen
  \bibfield  {author} {\bibinfo {author} {\bibfnamefont {L.~K.}\ \bibnamefont
  {Grover}},\ }in\ \href {\doibase 10.1145/237814.237866} {\emph {\bibinfo
  {booktitle} {Proceedings of the Twenty-eighth Annual ACM Symposium on Theory
  of Computing}}},\ \bibinfo {series and number} {STOC '96}\ (\bibinfo
  {publisher} {ACM},\ \bibinfo {address} {New York, NY, USA},\ \bibinfo {year}
  {1996})\ pp.\ \bibinfo {pages} {212--219}\BibitemShut {NoStop}%
\bibitem [{\citenamefont {Biham}\ \emph {et~al.}(2000)\citenamefont {Biham},
  \citenamefont {Biham}, \citenamefont {Biron}, \citenamefont {Grassl},
  \citenamefont {Lidar},\ and\ \citenamefont {Shapira}}]{Biham:2000aa}%
  \BibitemOpen
  \bibfield  {author} {\bibinfo {author} {\bibfnamefont {E.}~\bibnamefont
  {Biham}}, \bibinfo {author} {\bibfnamefont {O.}~\bibnamefont {Biham}},
  \bibinfo {author} {\bibfnamefont {D.}~\bibnamefont {Biron}}, \bibinfo
  {author} {\bibfnamefont {M.}~\bibnamefont {Grassl}}, \bibinfo {author}
  {\bibfnamefont {D.~A.}\ \bibnamefont {Lidar}}, \ and\ \bibinfo {author}
  {\bibfnamefont {D.}~\bibnamefont {Shapira}},\ }\href
  {http://link.aps.org/doi/10.1103/PhysRevA.63.012310} {\bibfield  {journal}
  {\bibinfo  {journal} {Physical Review A}\ }\textbf {\bibinfo {volume} {63}},\
  \bibinfo {pages} {012310} (\bibinfo {year} {2000})}\BibitemShut {NoStop}%
\bibitem [{\citenamefont {Biham}\ \emph {et~al.}(1999)\citenamefont {Biham},
  \citenamefont {Biham}, \citenamefont {Biron}, \citenamefont {Grassl},\ and\
  \citenamefont {Lidar}}]{Biham:1999ye}%
  \BibitemOpen
  \bibfield  {author} {\bibinfo {author} {\bibfnamefont {E.}~\bibnamefont
  {Biham}}, \bibinfo {author} {\bibfnamefont {O.}~\bibnamefont {Biham}},
  \bibinfo {author} {\bibfnamefont {D.}~\bibnamefont {Biron}}, \bibinfo
  {author} {\bibfnamefont {M.}~\bibnamefont {Grassl}}, \ and\ \bibinfo {author}
  {\bibfnamefont {D.~A.}\ \bibnamefont {Lidar}},\ }\href
  {http://link.aps.org/doi/10.1103/PhysRevA.60.2742} {\bibfield  {journal}
  {\bibinfo  {journal} {Physical Review A}\ }\textbf {\bibinfo {volume} {60}},\
  \bibinfo {pages} {2742} (\bibinfo {year} {1999})}\BibitemShut {NoStop}%
\bibitem [{\citenamefont {Schlissel}(1977)}]{schlissel1977initial}%
  \BibitemOpen
  \bibfield  {author} {\bibinfo {author} {\bibfnamefont {A.}~\bibnamefont
  {Schlissel}},\ }\href {https://doi.org/10.1016/0315-0860(77)90111-2}
  {\bibfield  {journal} {\bibinfo  {journal} {Historia Mathematica}\ }\textbf
  {\bibinfo {volume} {4}},\ \bibinfo {pages} {183} (\bibinfo {year}
  {1977})}\BibitemShut {NoStop}%
\bibitem [{\citenamefont {{A. Messiah}}(1999)}]{Messiah:book}%
  \BibitemOpen
  \bibfield  {author} {\bibinfo {author} {\bibnamefont {{A. Messiah}}},\
  }\href@noop {} {\emph {\bibinfo {title} {{Quantum Mechanics}}}}\ (\bibinfo
  {publisher} {{Dover Publication}},\ \bibinfo {address} {{New York}},\
  \bibinfo {year} {1999})\BibitemShut {NoStop}%
\bibitem [{\citenamefont {Teufel}(2003)}]{Teufel:book}%
  \BibitemOpen
  \bibfield  {author} {\bibinfo {author} {\bibfnamefont {S.}~\bibnamefont
  {Teufel}},\ }\href {http://link.springer.com/book/10.1007%2Fb13355} {\emph
  {\bibinfo {title} {Adiabatic Perturbation Theory in Quantum Dynamics}}},\
  \bibinfo {series} {Lecture Notes in Mathematics}, Vol.\ \bibinfo {volume}
  {1821}\ (\bibinfo  {publisher} {Springer-Verlag},\ \bibinfo {address}
  {Berlin},\ \bibinfo {year} {2003})\BibitemShut {NoStop}%
\bibitem [{\citenamefont {Hagedorn}\ and\ \citenamefont
  {Joye}(2002)}]{Hagedorn:2002kx}%
  \BibitemOpen
  \bibfield  {author} {\bibinfo {author} {\bibfnamefont {G.~A.}\ \bibnamefont
  {Hagedorn}}\ and\ \bibinfo {author} {\bibfnamefont {A.}~\bibnamefont
  {Joye}},\ }\href {\doibase 10.1006/jmaa.2001.7765} {\bibfield  {journal}
  {\bibinfo  {journal} {Journal of Mathematical Analysis and Applications}\
  }\textbf {\bibinfo {volume} {267}},\ \bibinfo {pages} {235} (\bibinfo {year}
  {2002})}\BibitemShut {NoStop}%
\bibitem [{\citenamefont {Holmes}(2012)}]{holmes2012introduction}%
  \BibitemOpen
  \bibfield  {author} {\bibinfo {author} {\bibfnamefont {M.~H.}\ \bibnamefont
  {Holmes}},\ }\href@noop {} {\emph {\bibinfo {title} {Introduction to
  perturbation methods}}},\ Vol.~\bibinfo {volume} {20}\ (\bibinfo  {publisher}
  {Springer Science \& Business Media},\ \bibinfo {year} {2012})\BibitemShut
  {NoStop}%
\bibitem [{\citenamefont {Winitzki}(2005)}]{Winitzki:2005aa}%
  \BibitemOpen
  \bibfield  {author} {\bibinfo {author} {\bibfnamefont {S.}~\bibnamefont
  {Winitzki}},\ }\href {http://link.aps.org/doi/10.1103/PhysRevD.72.104011}
  {\bibfield  {journal} {\bibinfo  {journal} {Physical Review D}\ }\textbf
  {\bibinfo {volume} {72}},\ \bibinfo {pages} {104011} (\bibinfo {year}
  {2005})}\BibitemShut {NoStop}%
\bibitem [{\citenamefont {Grover}(1997)}]{Grover:97a}%
  \BibitemOpen
  \bibfield  {author} {\bibinfo {author} {\bibfnamefont {L.~K.}\ \bibnamefont
  {Grover}},\ }\href {http://link.aps.org/doi/10.1103/PhysRevLett.79.325}
  {\bibfield  {journal} {\bibinfo  {journal} {Phys. Rev. Lett.}\ }\textbf
  {\bibinfo {volume} {79}},\ \bibinfo {pages} {325} (\bibinfo {year}
  {1997})}\BibitemShut {NoStop}%
\bibitem [{\citenamefont {Rezakhani}\ \emph {et~al.}(2010)\citenamefont
  {Rezakhani}, \citenamefont {Pimachev},\ and\ \citenamefont {Lidar}}]{RPL:10}%
  \BibitemOpen
  \bibfield  {author} {\bibinfo {author} {\bibfnamefont {A.~T.}\ \bibnamefont
  {Rezakhani}}, \bibinfo {author} {\bibfnamefont {A.~K.}\ \bibnamefont
  {Pimachev}}, \ and\ \bibinfo {author} {\bibfnamefont {D.~A.}\ \bibnamefont
  {Lidar}},\ }\href {http://link.aps.org/doi/10.1103/PhysRevA.82.052305}
  {\bibfield  {journal} {\bibinfo  {journal} {Phys. Rev. A}\ }\textbf {\bibinfo
  {volume} {82}},\ \bibinfo {pages} {052305} (\bibinfo {year}
  {2010})}\BibitemShut {NoStop}%
\bibitem [{\citenamefont {Bennett}\ \emph {et~al.}(1997)\citenamefont
  {Bennett}, \citenamefont {Bernstein}, \citenamefont {Brassard},\ and\
  \citenamefont {Vazirani}}]{Bennett:1997lh}%
  \BibitemOpen
  \bibfield  {author} {\bibinfo {author} {\bibfnamefont {C.}~\bibnamefont
  {Bennett}}, \bibinfo {author} {\bibfnamefont {E.}~\bibnamefont {Bernstein}},
  \bibinfo {author} {\bibfnamefont {G.}~\bibnamefont {Brassard}}, \ and\
  \bibinfo {author} {\bibfnamefont {U.}~\bibnamefont {Vazirani}},\ }\bibfield
  {booktitle} {\emph {\bibinfo {booktitle} {SIAM Journal on Computing}},\
  }\href {\doibase 10.1137/S0097539796300933} {\bibfield  {journal} {\bibinfo
  {journal} {SIAM Journal on Computing}\ }\textbf {\bibinfo {volume} {26}},\
  \bibinfo {pages} {1510} (\bibinfo {year} {1997})}\BibitemShut {NoStop}%
\bibitem [{\citenamefont {{A.P. Peirce, M.A. Dahleh, H.
  Rabitz}}(1988)}]{Peirce:88}%
  \BibitemOpen
  \bibfield  {author} {\bibinfo {author} {\bibnamefont {{A.P. Peirce, M.A.
  Dahleh, H. Rabitz}}},\ }\href {https://doi.org/10.1103/PhysRevA.37.4950}
  {\bibfield  {journal} {\bibinfo  {journal} {Phys. Rev. A}\ }\textbf {\bibinfo
  {volume} {37}},\ \bibinfo {pages} {4950} (\bibinfo {year}
  {1988})}\BibitemShut {NoStop}%
\bibitem [{\citenamefont {D'alessandro}\ and\ \citenamefont
  {Dahleh}(2001)}]{d2001optimal}%
  \BibitemOpen
  \bibfield  {author} {\bibinfo {author} {\bibfnamefont {D.}~\bibnamefont
  {D'alessandro}}\ and\ \bibinfo {author} {\bibfnamefont {M.}~\bibnamefont
  {Dahleh}},\ }\href {https://doi.org/10.1109/9.928587} {\bibfield  {journal}
  {\bibinfo  {journal} {IEEE Transactions on Automatic Control}\ }\textbf
  {\bibinfo {volume} {46}},\ \bibinfo {pages} {866} (\bibinfo {year}
  {2001})}\BibitemShut {NoStop}%
\bibitem [{\citenamefont {Grace}\ \emph {et~al.}(2007)\citenamefont {Grace},
  \citenamefont {Brif}, \citenamefont {Rabitz}, \citenamefont {Walmsley},
  \citenamefont {Kosut},\ and\ \citenamefont {Lidar}}]{Grace:2007aa}%
  \BibitemOpen
  \bibfield  {author} {\bibinfo {author} {\bibfnamefont {M.}~\bibnamefont
  {Grace}}, \bibinfo {author} {\bibfnamefont {C.}~\bibnamefont {Brif}},
  \bibinfo {author} {\bibfnamefont {H.}~\bibnamefont {Rabitz}}, \bibinfo
  {author} {\bibfnamefont {I.~A.}\ \bibnamefont {Walmsley}}, \bibinfo {author}
  {\bibfnamefont {R.~L.}\ \bibnamefont {Kosut}}, \ and\ \bibinfo {author}
  {\bibfnamefont {D.~A.}\ \bibnamefont {Lidar}},\ }\href
  {http://stacks.iop.org/0953-4075/40/i=9/a=S06} {\bibfield  {journal}
  {\bibinfo  {journal} {Journal of Physics B: Atomic, Molecular and Optical
  Physics}\ }\textbf {\bibinfo {volume} {40}},\ \bibinfo {pages} {S103}
  (\bibinfo {year} {2007})}\BibitemShut {NoStop}%
\bibitem [{\citenamefont {Barnes}\ and\ \citenamefont
  {Das~Sarma}(2012)}]{Barnes:2012fk}%
  \BibitemOpen
  \bibfield  {author} {\bibinfo {author} {\bibfnamefont {E.}~\bibnamefont
  {Barnes}}\ and\ \bibinfo {author} {\bibfnamefont {S.}~\bibnamefont
  {Das~Sarma}},\ }\href
  {http://link.aps.org/doi/10.1103/PhysRevLett.109.060401} {\bibfield
  {journal} {\bibinfo  {journal} {Physical Review Letters}\ }\textbf {\bibinfo
  {volume} {109}},\ \bibinfo {pages} {060401} (\bibinfo {year}
  {2012})}\BibitemShut {NoStop}%
\bibitem [{\citenamefont {Palittapongarnpim}\ \emph {et~al.}(2017)\citenamefont
  {Palittapongarnpim}, \citenamefont {Wittek}, \citenamefont {Zahedinejad},
  \citenamefont {Vedaie},\ and\ \citenamefont
  {Sanders}}]{palittapongarnpim2017learning}%
  \BibitemOpen
  \bibfield  {author} {\bibinfo {author} {\bibfnamefont {P.}~\bibnamefont
  {Palittapongarnpim}}, \bibinfo {author} {\bibfnamefont {P.}~\bibnamefont
  {Wittek}}, \bibinfo {author} {\bibfnamefont {E.}~\bibnamefont {Zahedinejad}},
  \bibinfo {author} {\bibfnamefont {S.}~\bibnamefont {Vedaie}}, \ and\ \bibinfo
  {author} {\bibfnamefont {B.~C.}\ \bibnamefont {Sanders}},\ }\href
  {https://doi.org/10.1016/j.neucom.2016.12.087} {\bibfield  {journal}
  {\bibinfo  {journal} {Neurocomputing}\ } (\bibinfo {year}
  {2017})}\BibitemShut {NoStop}%
\bibitem [{\citenamefont {Nenciu}(1993)}]{Nenciu:93}%
  \BibitemOpen
  \bibfield  {author} {\bibinfo {author} {\bibfnamefont {G.}~\bibnamefont
  {Nenciu}},\ }\href {\doibase 10.1007/BF02096616} {\bibfield  {journal}
  {\bibinfo  {journal} {Commun.Math. Phys.}\ }\textbf {\bibinfo {volume}
  {152}},\ \bibinfo {pages} {479} (\bibinfo {year} {1993})}\BibitemShut
  {NoStop}%
\bibitem [{\citenamefont {Lidar}\ \emph {et~al.}(2009)\citenamefont {Lidar},
  \citenamefont {Rezakhani},\ and\ \citenamefont {Hamma}}]{lidar:102106}%
  \BibitemOpen
  \bibfield  {author} {\bibinfo {author} {\bibfnamefont {D.~A.}\ \bibnamefont
  {Lidar}}, \bibinfo {author} {\bibfnamefont {A.~T.}\ \bibnamefont
  {Rezakhani}}, \ and\ \bibinfo {author} {\bibfnamefont {A.}~\bibnamefont
  {Hamma}},\ }\href
  {http://scitation.aip.org/content/aip/journal/jmp/50/10/10.1063/1.3236685}
  {\bibfield  {journal} {\bibinfo  {journal} {J. Math. Phys.}\ }\textbf
  {\bibinfo {volume} {50}},\ \bibinfo {pages} {102106} (\bibinfo {year}
  {2009})}\BibitemShut {NoStop}%
\bibitem [{\citenamefont {Ge}\ \emph {et~al.}(2016)\citenamefont {Ge},
  \citenamefont {Moln{\'a}r},\ and\ \citenamefont {Cirac}}]{Ge:2015wo}%
  \BibitemOpen
  \bibfield  {author} {\bibinfo {author} {\bibfnamefont {Y.}~\bibnamefont
  {Ge}}, \bibinfo {author} {\bibfnamefont {A.}~\bibnamefont {Moln{\'a}r}}, \
  and\ \bibinfo {author} {\bibfnamefont {J.~I.}\ \bibnamefont {Cirac}},\ }\href
  {http://link.aps.org/doi/10.1103/PhysRevLett.116.080503} {\bibfield
  {journal} {\bibinfo  {journal} {Physical Review Letters}\ }\textbf {\bibinfo
  {volume} {116}},\ \bibinfo {pages} {080503} (\bibinfo {year}
  {2016})}\BibitemShut {NoStop}%
\end{thebibliography}%

%%%%%%%%%%%%%%
%%%%%%%%%%%%%%%%
%%%%%%%%%%%%%%%%%%%

\appendix

%\onecolumngrid

%%%%

%%%%

%%%%

%%%%%%%%%%%%%
\section{Comparison with the method of Hagedorn and Joye} \label{app:Hage}

In this section, we recap the asymptotic expansion of Hagedorn and Joye~\cite{Hagedorn:2002kx}, which is a powerful tool for proving adiabatic theorems. In particular, the Hagedorn and Joye method can be used to prove bounds on the error incurred due to their asymptotic expansion; in fact, the main goal of Ref.~\cite{Hagedorn:2002kx} was to show that the adiabatic approximation can provide exponentially small errors if the Hamiltonian is analytic in the time-variable (see also Refs.~\cite{Nenciu:93,lidar:102106,Ge:2015wo}). Here, we analyze its utility as a computational tool. 

Hagedorn and Joye (HJ) propose the following method to obtain asymptotic approximations to the time-dependent Schr{\"o}dinger equation
\beq
i \epsilon \frac{d \ket{\chi(r)}}{dr} = \mathcal{H}(r) \ket{\chi(r)}.
\eeq
Note that the above equation is of the form of Eq.~\eqref{eq:rescaleSchro}, with $\epsilon \equiv \frac{1}{\mu t_f}$ and $\mathcal{H}(r) \equiv s^\pr(r) H(r)$.

They obtain a theorem which states that for any value of the small parameter $\epsilon$, one can write down an approximation for $\ket{\chi(r)}$ which takes the form of a power series in $\epsilon$. The quality of the approximation (as measured by the 2-norm) scales as $e^{-\frac{1}{\epsilon}}$ provided that the number of terms in the series scales as $1/\epsilon$. More precisely:
\begin{theorem}[\cite{Hagedorn:2002kx}] Assume reasonable smoothness and gap conditions on the Hamiltonian. We can then recursively obtain an asymptotic expansion of the form
\begin{multline}
\ket{\chi_\mathrm{HJ}^{(N)} (r,\epsilon)} = e^{-\frac{i}{\epsilon} \int_0^r E(q) dq} \left(\ket{\chi_0(r)} + \epsilon \ket{\chi_1(r)} \right. \\ 
\left. + \dots + \epsilon^N \ket{\chi_N (r)} + \epsilon^{N+1} \ket{\chi_{N+1}^\perp (r)} \right).
\end{multline}
such that for any $r$, there exist positive $G$, $C(g)$, and $\Gamma(g)$ such that for all $g \in (0,G)$, the vector $\ket{\chi^{(\lfloor{g/\epsilon}\rfloor)}_\mathrm{HJ} (r,\epsilon)}$ satisfies
\beq
\| \ket{\chi(r,\epsilon)} - \ket{\chi^{(\lfloor{g/\epsilon}\rfloor)}_\mathrm{HJ} (r,\epsilon)} \|_2 \leq C(g) e^{-\Gamma(g)/\epsilon},
\eeq
for all $\epsilon \leq 1$. Here, $\ket{\chi(r,\epsilon)}$ is the Schr{\"o}dinger evolved wavefunction starting from the initial condition $\ket{\chi(0,\epsilon)} =\ket{\chi^{(\lfloor{g/\epsilon}\rfloor)}_\mathrm{HJ} (0,\epsilon)}$. 
\end{theorem}

We explore the usefulness of this asymptotic expansion as an approximation tool and thus we do not estimate the number of terms that are necessary to provide an exponentially small error. Instead, we develop the approximation for two orders and compare the resulting asymptotic expansion with the WKB method.

Let us develop the terms in the HJ expansion (as given in Ref.~\cite{Hagedorn:2002kx}). We substitute the asymptotic ansatz
\beq \label{eq:HJasympt}
\ket{\chi_\mathrm{HJ}} \sim e^{-\frac{i}{\epsilon} \int_0^r dq E(q)} \left( \ket{\chi_0(r)} + \epsilon \ket{\chi_1(r)} + \dots\right)
\eeq
into the Schr{\"o}dinger equation, and equate the terms multiplying the same order of $\epsilon$, which results in the following expression for the $O(\epsilon^j)$ term
\beq
\ket{\chi_j (r)} = f_j(r) \ket{\Phi(r)} + \ket{\chi^\perp_j (r)}.
\eeq
Here $\ket{\Phi(r)}$ is the eigenstate being (approximately) followed, and the other components of the above formula are obtained recursively by using:
\bes
\begin{align}
f_0(r) &= 1; \\
f_{j-1} (r) &= -\int_0^r \braket{\Phi(q) | \partial_q \psi_{j-1}^\perp (q)} dq, \quad j\geq 2 \label{eqt:before}\\
&=\int_0^r \braket{\Phi^\pr(q) | \psi_{j-1}^\perp (q)} dq \label{eqt:after};\\
\ket{\chi_j^\perp (r)} &= i [H(r)-E(r)]_R^{-1} \left( f_{n-1}(r) \ket{\Phi^\pr (r)} \right. \nonumber \\
& \qquad \left. + P_\perp(r)\partial_r \ket{\chi_{j-1}^\perp (r)} \right);
\end{align}
\ees
where, in going from Eq.~\eqref{eqt:before} to Eq.~\eqref{eqt:after}, we integrated by parts and used $\braket{\Phi(q)|\psi_{j-1}^\perp(q)}=0$. Also, $P_\perp(r) \equiv I - \ket{\Phi(r)}\bra{\Phi(r)}$ is the instantaneous projector on to the complement of $\ket{\Phi(r)}$; $E(r)$ is the eigenvalue being quasi-adiabatically followed; and $[H(r)-E(r)]_R^{-1}$ is the reduced resolvent, i.e., the inverse of $[H(r)-E(r)]$ restricted to the complement of $\ket{\Phi(r)}$.

In order to compare the HJ expansion with the WKB approximation, we will compare the $N$-th order expansion provided by both methods. Note that the $N$-th order of the HJ expansion includes terms up to $\mathcal{O}(\epsilon^{N+1})$. 
This means that we will be comparing the zeroth order of WKB (i.e., $\ket{\chi_\mathrm{WKB}^{(0)}}$) with 
\beq \label{eq:HJord1}
\ket{\chi_\mathrm{HJ}^{(0)}(r)} \equiv e^{-\frac{i}{\epsilon} \int_0^r E(q) dq} \left(\ket{\chi_0(r)} + \epsilon \ket{\chi^\perp_1(r)} \right);
\eeq
and the first order WKB (i.e., $\ket{\chi_\mathrm{WKB}^{(1)}}$) with 
\beq
\ket{\chi_\mathrm{HJ}^{(1)}(r)} \equiv e^{-\frac{i}{\epsilon} \int_0^r E(q) dq} \left(\ket{\chi_0(r)} + \epsilon \ket{\chi_1(r)} + \epsilon^2 \ket{\chi^\perp_2(r)}\right).
\eeq

For two-level systems such as the one that we are concerned with, we obtain the following simplified expressions, where ``$\mathrm{GS}$" and ``$\mathrm{Exc}$" denote the ground and excited states respectively and $\Delta$ represents the spectral gap:
\bes
\begin{align}
[H(r)-E_\mathrm{GS}(r) ]_R^{-1} &= \frac{1}{\Delta(r)} \ket{\chi_\mathrm{Exc}(r)}\bra{\chi_\mathrm{Exc}(r)} \\
\implies \quad \ket{\chi_1^\perp(r)} &= \frac{i}{\Delta(r)} \braket{\chi_\mathrm{Exc}(r) | \chi_\mathrm{GS}^\pr(r)}\ket{\chi_\mathrm{Exc}(r)}\ , \\
f_1(r) &= \int_0^r dq \braket{\chi^\pr_\mathrm{GS}(q) | \chi_1^\perp(q)}\ ,  \\
\ket{\chi_2^\perp(r)} &= \frac{i}{\Delta(r)}  \left( f_1(r) \ket{\chi_\mathrm{GS}^\pr(r)} \right.  \\
& \left. +  \ket{\chi_\mathrm{Exc}(r)}\braket{ \chi_\mathrm{Exc}(r) | \partial_r\chi_1^\perp(r)} \right) \nonumber\ .
\end{align}
\ees
We have assumed that the ground state is being followed and hence set $\ket{\Phi} = \ket{\chi_\mathrm{GS}}$. We have also used the fact that for real-valued Hamiltonians in two dimensions $\braket{\chi_\mathrm{GS}^\pr | \chi_\mathrm{GS}} = 0$. (Note that this does not mean $\ket{\chi_\mathrm{GS}^\pr} = \ket{\chi_\mathrm{Exc}}$ because $\ket{\chi_\mathrm{GS}^\pr}$ is generally not normalized and carries a non-trivial phase.)

\begin{figure*}[t]
\hspace*{\fill}
\subfigure[\ ]{\includegraphics[width = \columnwidth]{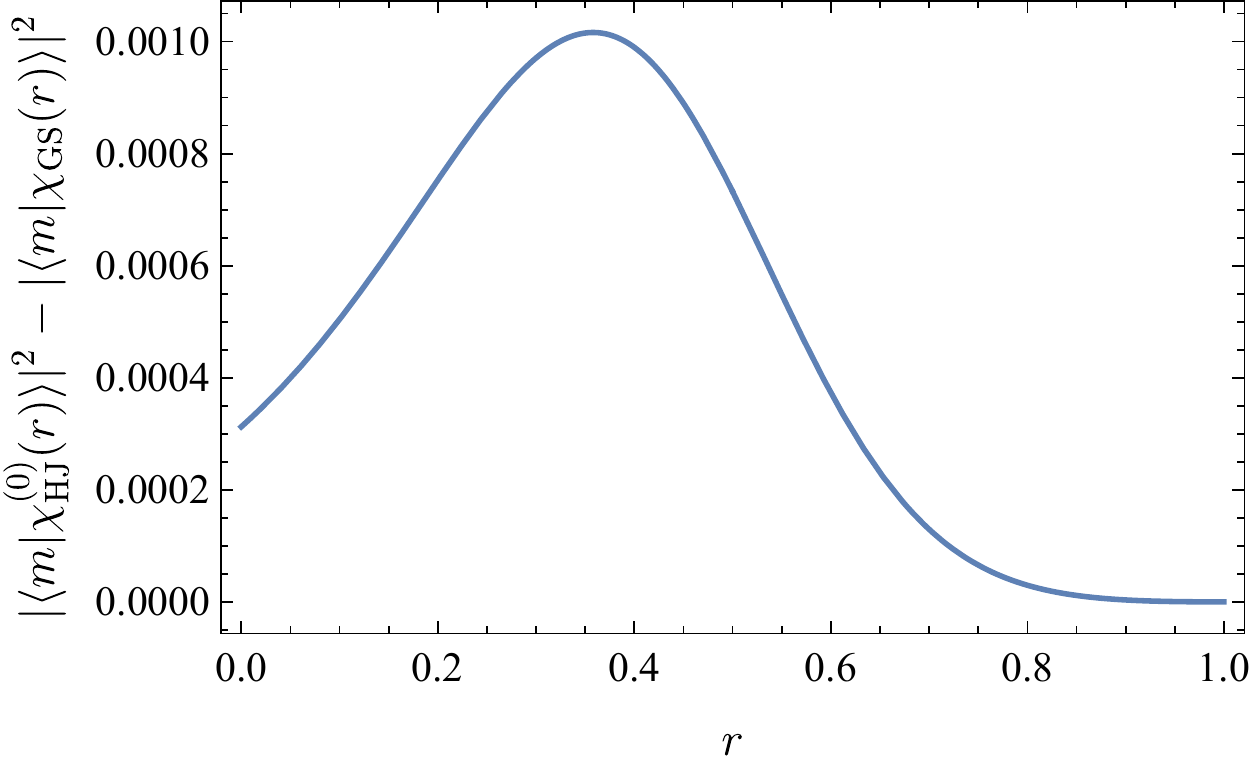}\label{fig:adiabaticHJ0popdiff}} \hfill\hfill
\subfigure[\ ]{\includegraphics[width = \columnwidth]{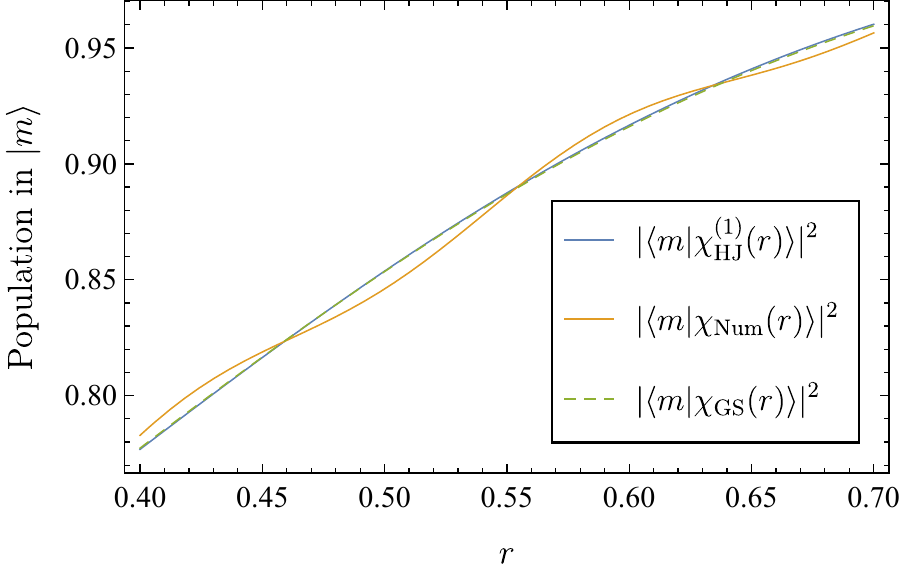}\label{fig:HJ2orderNumAdiak=1}}
\hspace*{\fill}
\caption{(a) The difference between the predictions of the naive adiabatic approximation ($\ket{\chi_\mathrm{GS}(r)}$) and the lowest order HJ approximation ($\ket{\chi_\mathrm{HJ}^{(0)}}$) for the population in the state $\ket{\chi_\mathrm{GS}(1)}\equiv\ket{0}$, as a function of the rescaled time parameter $r$, for $t_f=20$. The difference is very small, of the order of $10^{-3}$. (b) The population in the state $\ket{m}$ as a function of time for the HJ expansion using the $0$th and $1$st orders; the adiabatic solution; and the numerical solution. The adiabatic solution and the HJ method are indistinguishable on the scale of this plot. Clearly, they do not the capture the oscillations displayed by the numerical solution. Here $t_f=50$.}
\end{figure*}

We now restrict to the case of a qubit in a magnetic field. 

First, consider $\ket{\chi_\mathrm{HJ}^{(0)}}$ (which includes terms up to order $\epsilon$). Figure~\ref{fig:adiabaticHJ0popdiff} shows that $\ket{\chi_\mathrm{HJ}^{(0)}}$ provides an approximation that is `too adiabatic'. In particular, it fails to capture the oscillations that are captured by the WKB approximation, as seen in Fig.~\ref{fig:popDynamics_n=1}. Furthermore, from the form of $\ket{\chi_\mathrm{HJ}^{(0)}}$ it is clear that this approximation will predict $p_\mathrm{GS}(t_f) = 1$ always:
\begin{align}
&p_\mathrm{GS}^{\mathrm{HJ},0}(t_f) = \abs{\braket{\chi_\mathrm{GS}(1) | \chi_\mathrm{HJ}^{(0)}}}^2 \\
&= |\underbrace{\braket{\chi_\mathrm{GS}(1)|\chi_\mathrm{GS}(1)}}_{=1} + \underbrace{\braket{\chi_\mathrm{GS}(1)|\chi_1^\perp(1)}}_{=0}|^2.
\end{align}

Next, consider $\ket{\chi_\mathrm{HJ}^{(1)}}$ (which includes terms up to order $\epsilon^2$). Figure~\ref{fig:HJ2orderNumAdiak=1} shows that this too provides an approximation which fails to capture the  oscillations that are present in the numerical solution and also in the lowest order WKB solution. Thus, we conclude that the WKB method is more suitable for developing analytic approximations.

%Why does the WKB do better? Recall that the WKB approach requires us to decouple the solutions for $\psi$ and $\phi$ (see the end of Sec.~\ref{sec:WKBmethod}). This is a crucial distinction between our method and the HJ method. We speculate that it is this ``decoupled" aspect of the WKB method (while more computationally intensive) that allows for a better approximation performance compared to the ``coupled" HJ method. \DL{Can you offer any evidence beyond speculation for why this is the crucial point? If not, it's better to drop this paragraph.}

While we pointed out some of the disadvantages of the HJ method as an approximation technique, we remark that the method is particularly useful to prove scaling results. For example, consider,
\begin{align}
\abs{\braket{\chi_\mathrm{GS}(1)|\chi_\mathrm{HJ}^1(1)}}^2 &= \abs{\left(1+ \epsilon f_1(1)\right)}^2 \\
&=  \left(1+ \epsilon^2\abs{f_1(1)}^2\right)\\
&= \mathcal{O}(1) + \epsilon^2 \mathcal{O}(1).
\end{align}
In the first line, we used the fact $\ket{\chi_\mathrm{GS}}$ is orthogonal to any (unnormalized) state that carries the $\perp$ symbol. In the second line, we  used the fact that 
\beq
f_1(1) = i \int_0^1 dq \frac{\abs{\braket{\chi_\mathrm{GS}^\pr(q)|\chi_\mathrm{Exc}(q)}}^2}{\Delta(q)}
\eeq
is purely imaginary. Thus the HJ expansion captures the $1-\bigo(\frac{1}{t_f^2})$ scaling of the final ground state probability. 

%%%%%

%\section{$p_\mathrm{GS}$ vs. $t_f$ curves for different schedules} \label{app:pgs}

%In this Appendix we show the $p_\mathrm{GS}(t_f)$ vs. $t_f$ curves for the four different schedules, $g_\alpha$, $\alpha \in \{1,2,3,4\}$. 

%We see that for the $g_0$ and $g_1$ schedules, the probability quickly rises above $1$. Thus, these schedules quickly result in unphysical results. We hypothesize that this is the reason why these schedules result in $t_\mathrm{Th}(n)$ scalings that violate the query complexity bound. We further see that the rise in $p_\mathrm{GS}$ with $t_f$ is faster for lower $\alpha$. Thus, somehow, the $g_2$ schedule captures the right scaling by capturing the right steepness: i.e., for $\alpha < 2$, the rise is too steep; and for $\alpha > 2$, the rise is too shallow.

%%%%%%%%%%%%%%%%%

%\begin{figure}[!htbp]
%\centering
%\includegraphics[width = 0.9\columnwidth]{IntOneNormTrd_vs_tf_n=6_renormedWKB0}
%\caption{The time-averaged trace-norm distance [Eq.~\eqref{eqt:inttrd}] between the renormalized WKB approximation at the lowest order and the numerically exact solution for the four different schedules [Eqs.~\eqref{eq:4schedsprime}]. $g_2$ represents the optimal schedule found in Ref.~\cite{Roland:2002ul}, which provides the best approximation to the numerical evolution. \red{SM: How is this related to Fig.~\ref{fig:manyschedsrenormedWKB}?}. Here $n=6$.}
%\label{fig:manyschedsITDrenormedWKBNum}
%\end{figure}

\end{document}